\documentclass{aa}  

\usepackage{graphicx}
\usepackage{siunitx}
\usepackage{txfonts}
\usepackage{lipsum}
\usepackage{subcaption}
\usepackage{lscape}
\usepackage{placeins}

\begin{document}

   \title{Tracing Magnetic Helicity in Interplanetary Space}

   \subtitle{A Multi-Spacecraft Analysis of the 2024 March 23 ICME}

   \author{A. Moulin\inst{1}, J.K. Thalmann\inst{1}, A.M. Veronig \inst{1,2}, M. Dumbovi\'{c}\inst{3}, H.T. R\"{u}disser\inst{1,4}, \\E.E. Davies\inst{4}, C. M\"{o}stl\inst{4}
        \and U.V. Amerstorfer\inst{4}
        }

   \institute{University of Graz, Institute of Physics, Universitätsplatz 5, 8010 Graz, Austria\\
             \email{anthony.moulin@uni-graz.at}
            \and University of Graz, Kanzelhöhe Observatory for Solar and Environmental Research, Kanzelhöhe 19, 9521 Treffen, Austria
            \and Hvar Observatory, Faculty of Geodesy, University of Zagreb, Kaciceva 26, 10000 Zagreb, Croatia
            \and Austrian Space Weather Office, GeoSphere Austria, Reininghausstr. 3, 8020 Graz, Austria}

   \date{Received September 30, 20XX}

  \abstract 
   {Probing interplanetary coronal mass ejections (ICMEs) and, more specifically, their magnetic field structure during their propagation in the heliosphere is critical to understanding the impacts they might have at 1 AU. Notably, a key quantity linking different spatial regimes is magnetic helicity, as it is a quasi-conserved value in resistive magnetohydrodynamics.}
   {We study the evolution of the magnetic helicity content of an ICME event by inferring its magnetic-field-related quantities at different stages between the Sun and Earth using remote sensing and in-situ observations.}
   {We study an eruptive X1.1 flare/CME event and its associated ICME originating from NOAA active region (AR) 13614 on 2024 March 23 00:58 UT. Using magnetogram data of the AR during the flaring event, along with non-linear force-free (NLFF) magnetic field modeling, we reconstruct the magnetic field in the solar corona and analyze its magnetic helicity budget. The helicity is also tracked in-situ in the heliosphere using four well radially-aligned spacecraft (Solar Orbiter at 0.39 AU, BepiColombo at 0.58 AU, STEREO-A at 0.96 AU and Wind at 0.99 AU) measuring the ICME magnetic field signature. We use different force-free magnetic field models (Lundquist, Gold-Hoyle) to evaluate the helicity values at each spacecraft. With the Gold-Hoyle-based model (expansion-modified force-free model, 3DCORE), we also reconstruct the ICME and its propagation in the heliosphere. We ultimately compare the results obtained for each model and spacecraft to the values computed in the corona to give a general understanding of how the event's magnetic field evolved.}
   {The NLFF modeling of the AR indicates a left-handed (negative helicity) flux rope (FR) positioned along a slightly northeast-southwest axis, in agreement with chromospheric and coronal observations (H-$\alpha$, SDO/AIA), and yields a helicity budget computed of $-2.16\pm1.32\times10^{41}~\text{Mx}^2$. The helicity inferred from in-situ modeling methods reveals values in a similar range ($-2.0\pm1.0\times10^{41}~\text{Mx}^2$), but is highly sensitive to the spacecraft crossing position through the ICME. It appears that the results given by 3DCORE are more consistent between all spacecraft, as well as with the values inferred at the AR.}
   {}

    \authorrunning{Moulin et al.}
   \maketitle
   \nolinenumbers

\section{Introduction}
Coronal mass ejections (CMEs) are powerful expulsions of magnetized coronal plasma from the solar atmosphere with speeds of hundreds to thousands of kilometers per second. Their large-scale propagation and expansion into the heliosphere are known as interplanetary CMEs \citep[ICMEs;][]{kilpua_coronal_2017} and are distinct from the solar wind because of their magnetic field strength, declining velocity profile, low plasma $\beta$ and other features \citep{bothmer_structure_1998, zurbuchen_-situ_2006}. An Earth-directed ICME can severely affect our space weather and thus our technology on ground and in space \citep[e.g.][]{schwenn_space_2006, temmer_space_2021, cliver_extreme_2022}. ICME properties such as magnetic field orientation are often measured in-situ by spacecraft at different positions in the heliosphere. These measurements helped develop a representation of ICMEs where the front shock is often followed by a sheath region characterized by a turbulent and compressed magnetic field. The core of the ICME is located in the following part where the magnetic field is enhanced and smoothly rotating, indicative of a magnetic flux rope (MFR), defined by helical field lines coiled around a common central axis. ICME magnetic field structures with clear flux rope features are often described as magnetic clouds (MCs; \citet{burlaga_magnetic_1981}).\\
\indent Previous studies have depicted a classical axi-symmetric geometry \citep{zurbuchen_-situ_2006, wang_observations_2018} and a self-similar expansion in the radial direction \citep{demoulin_causes_2009, vrsnak_heliospheric_2019} that a large part of observed ICMEs exhibit \citep{nieves-chinchilla_unraveling_2019, cane_interplanetary_2003}. More recently, this understanding is evolving towards a truly 3-dimensional structure with non-uniform cross-section and twist \citep{al-haddad_magnetic_2025}. Their expansion is also believed to be significantly impacted by the background solar wind velocity and the magnetic pressure difference \citep{zhang_studying_2025}.\\
\indent ICMEs increase in size and decrease in magnetic field strength during their propagation can be modeled as power-law functions \citep[e.g.][]{davies_flux_2024, salman_survey_2024, leitner_consequences_2007}. The magnetic field decay index ($n_b$) best fit was shown to be of the order of $n_b = 1.6\pm0.2$ by \citet{patsourakos_near-sun_2016} and more recently by \citet{mostl_magnetic_2026}, whereas the expansion factor was constrained to $0.45<n_a<1.14$ \citep{gulisano_expansion_2012, davies_multi-spacecraft_2022}. Several phenomena can impact ICME propagation in the heliosphere and explain the observed differences in in-situ magnetic field measurements, even with low spacecraft separation \citep{lugaz_spatial_2018, davies_radial_2020, regnault_discrepancies_2024}. Different sources may cause these differences, such as magnetic configuration, solar wind drag, as well as reconnection processes between the flux rope and the sheath region \citep{luhmann_icme_2020}. The magnetic reconnection between the ejecta and sheath region has been interpreted as an ``erosion'' of the flux rope \citep{ruffenach_statistical_2015, wang_observations_2018}, changing its geoeffectiveness. Moreover, CMEs that occur in close succession can interact with each other if the following one is faster than the first one, causing great changes in the CME in-situ profile between the Sun and Earth. Their interaction is a complex phenomenon (depending on their respective direction of propagation, speed, magnetic field orientation and timing) which is not yet fully understood \citep{sabri_quantifying_2026, manchester_physical_2017}. Additionally, the flux rope shape is evolving during its propagation in the heliosphere, leading to a flattening of the cross section referred to as ``pancaking effect'' \citep{riley_kinematic_2004, davies_situ_2021}.\\
\indent Most of the CME/ICME events originate from solar active regions and are associated with solar flares \citep{webb_coronal_2012}. Energetic flares, especially X-class flares, have a CME-association rate of about 90\% \citep{yashiro_different_2006}. A large eruptive flare also unveils growing magnetic field arcades (visible in soft X-ray (SXR) and extreme ultra-violet (EUV) imaging) exhibiting their footprint in the lower solar atmosphere, often in the form of two bands of bright emission (so-called flare ribbons) on either side of the polarity inversion line \citep{fletcher_observational_2011, benz_flare_2017}. CMEs erupting from a flaring active region are also often associated with coronal dimmings, i.e. a sudden decrease of coronal emission in EUV and SXR. They are interpreted as "footprints" of the erupting flux rope and are directly linked to the mass loss. For details on flaring ARs and coronal dimmings, we refer to the reviews by \citet{toriumi_flare-productive_2019} and \citet{veronig_coronal_2025}.\\
\indent A common way to investigate the solar counterpart of an ICME is to study the magnetic helicity of the source active region, which is a measure of the shear or twist of the magnetic field. An important property of helicity is its conservation (quasi-conservation) in ideal (resistive) magnetohydrodynamics \citep{Berger&Field_1984}, because its dissipation time is longer than the energy dissipation time. This conservation suggests that the helicity budget of a pre-eruptive active region should be related to the helicity content in the MC ejected from the flaring AR, provided that the magnetic complexity of the ejected structure is not altered by processes during its travel in interplanetary space  \citep{lugaz_interaction_2013}. Such a comparison can be done by evaluating helicity using ICME magnetic field signatures measured by different spacecraft in the heliosphere \citep{zhang_studying_2025}. By fitting these signatures with a model assuming a basic structure of the field and topology, it is possible to infer the magnetic helicity content of the flux rope and then compare it to the helicity budget of the source active region. A few studies have attempted such comparisons \citep{dumbovic_2019_2021, thalmann_tracking_2023, al-haddad_magnetic_2025}, and advocate for comparisons to in-situ measurements with spacecraft closer to the Sun, to reduce the changes that may happen during the ICME propagation in interplanetary space.\\
\indent The increasing number of spacecraft in the inner heliosphere, notably with the addition of Solar Orbiter (SolO) and Parker Solar Probe (PSP), may provide this needed data. Combining them with L1 measurements allows multiple spacecraft to probe the same MC. This multi-regime (from Sun to Earth) approach can also show evidence of different processes altering the ICME helicity content. We present here the analysis of an ICME detected by a constellation of four spacecraft (i.e., Solar Orbiter, BepiColombo, STEREO-A and Wind) between 2024 March 23 and 2024 March 26 at different radial distances from the Sun, from 0.39 AU up to \textasciitilde{}1 AU. The ICMEs probed by these spacecraft are associated with this flare by the lineup catalog available online\footnote{https://helioforecast.space/lineups} \citep[see][]{mostl_multipoint_2022}. This configuration enables the tracking of key physical parameters from the Sun to Solar Orbiter and onward to near-Earth orbit. The main objective of this study is to study the magnetic helicity evolution of this ICME, and to connect it to the helicity of the solar source region and the related X1.1 flare/CME event.
\section{Data and Methods}
\subsection{Remote sensing analysis}\label{ss:RemoteSensing}
The flare/CME event is erupting from the active region NOAA AR3614 and is observed by the Atmospheric Imaging Assembly \citep[AIA;][]{lemen_atmospheric_2012} on board the Solar Dynamic Observatory \citep[SDO;][]{pesnell_solar_2012}, allowing us to analyze the coronal evolution related to the CME occurrence. Combining these observations with the National Solar Observatory Global Oscillation Network Group (NSO GONG) H-$\alpha$ imaging, we can also understand the low-atmosphere imprint of the CME-associated solar flare (SOL2024-03-23T01:33X1.1).\\
\indent To study the magnetic field structure and helicity in detail during the event, we use the Helioseismic and Magnetic Imager (HMI; \citet{schou_design_2012, scherrer_helioseismic_2012}) on board SDO. It provides vector magnetic field measurements using a set of 12 observations from the Sun taken at different wavelengths around 6173.3 $\AA$. In particular, we used {\sc hmi.sharp\_cea\_720s} data, which provides the Lambert cylindrical equal area (CEA) projected heliographic magnetic field vector, within automatically identified AR Patches (SHARP; \citet{bobra_helioseismic_2014}), with a spatial sampling of 0.03 CEA-degrees (corresponding to $\sim$\,360\,km at disk center) and a time cadence of 12~minutes. We study the time range 2024 March 22 20:00 UT - 2024 March 23 05:24 UT during the flare/CME event, which is located around N25E05. The tangential ($Bx$, $By$) and normal ($Bz$) components of the measured  magnetic field vector were used as input to non-linear force-free (NLFF) modeling.\\
\indent For completeness, we also acknowledge the presence of another flare event originating from the active region NOAA AR3615 located around S05W05, and erupting $\sim$30 minutes after the X1.1 flare/CME under study. Even though this flare could impact the class attributed to the other eruption (due to GOES full-Sun flux measurements), the slow speed of the eruption, the absence of clear flux rope counterparts visible with remote sensing imaging, and the in-situ flux rope characteristics indicate that there is no CME-CME interaction between those two events. \\
\indent It is possible to use the vector magnetic field magnetogram data to provide an estimate of the three-dimensional distribution of the coronal magnetic field, which is fundamental to infer the helicity budget. Under coronal equilibrium conditions, non-linear force-free models are the more realistic models \citep{wiegelmann_solar_2021}. These models solve, inside the simulation volume, the differential equation system:
\begin{gather}
    (\nabla\times\boldsymbol{B})\times\boldsymbol{B} = 0\label{eqs:NLFFsystem1}\\
    \nabla\cdot\boldsymbol{B} = 0\label{eqs:NLFFsystem2}
\end{gather}

\noindent where $\boldsymbol{B}$ is the 3D coronal magnetic field above the measured surface field (boundary condition).\\
\indent The NLFF coronal magnetic field extrapolation is made using a physics-informed neural network \cite[NF2,][]{jarolim_probing_2023}. The NF2 code is very efficient in extrapolating series of vector magnetic fields, as it computes a full extrapolation of the first frame and uses the results as a starting condition for the next one. From a technical point of view, NF2 finds a trade-off between the physical model and the measurements through the setting of the divergence-free ($\lambda_{div})$ and force-free ($\lambda_{ff}$) parameters. As the quality of the extrapolation is important to compute the magnetic helicity evolution, different sets of parameters were tested.\\
\indent Assessing the level of solenoidality of the solution is thus important for helicity computation \citep{valori_accuracy_2013}. We compare the quality of the extrapolations by computing two values estimating their force- and divergence-freeness (Eqs.\ref{eqs:NLFFsystem1}, \ref{eqs:NLFFsystem2}). The first one is important to quantify how well the system has been solved for a Lorentz force equal to $0$, and the second has been proven important for the reliability of this solution \citep{valori_magnetic_2016}. The force-freeness is evaluated using the current-weighted angle ($\theta_J$) between the modeled magnetic field and the current electric density. An ideal (completely force-free) value is $\theta_J = 0^\circ$. The second criterion, $E_{div}/E$, assesses the level of solenoidality of the extrapolated magnetic field by defining the non-solenoidal fraction of the total energy \citep{valori_accuracy_2013}. The value of $E_{div}$ comes from the decomposition of the potential ($B_0$) and current-carrying ($B_J$) magnetic field into solenoidal and non-solenoidal parts. The $E_{div}/E$ threshold to obtain reliable magnetic helicity computation was originally thought to be around $E_{div}/E \lessapprox 0.08$, but \citet{thalmann_reliability_2019} proposed an even lower value, $E_{div}/E \lessapprox 0.05$.\\
\indent From the reliable extrapolations, we infer the magnetic helicity and its current-carrying component by using a method from \citet{thalmann_estimating_2011}. This method computes the vector potential from the given magnetic field in a volume $V$. If we consider $V$ as magnetically closed (i.e., $\boldsymbol{n\cdot B} = B_n = 0$, $n$ being the normal vector pointing in a normal outward direction), helicity can then be written as
\begin{equation}
    H_V = \int_V \boldsymbol{A\cdot B}\hspace{0.2cm}dV
\end{equation}
where $\boldsymbol{A}$ denotes the magnetic vector potential and $\boldsymbol{B} = \nabla \times \boldsymbol{A}$ is the solenoidal magnetic field contained in $V$. This formulation of helicity is not candidly applicable to the solar corona case, as the volume considered is not gauge independent (magnetic flux is constantly penetrating through the solar photosphere). Numerous solutions to this problem have been proposed \citep{demoulin_recent_2007}. \citet{Berger&Field_1984} and \citet{Finn&Antonsen_1984} introduced the relative helicity as:
\begin{equation}\label{eq:relH}
    H_V = \int_V (\boldsymbol{A}+\boldsymbol{A_0})\cdot(\boldsymbol{B}-\boldsymbol{B_0})\hspace{0.2cm}dV
\end{equation}
where $\boldsymbol{B_0} = \nabla\times \boldsymbol{A_0}$ is a reference field (usually chosen as a potential field) and its associated vector potential. $\boldsymbol{B}$ and $\boldsymbol{B_0}$ are sharing their normal component with $V$.\\
\indent This gauge-invariant quantity can be split into two different, also gauge-invariant ones \citep{berger_introduction_1999} in the form $H_V = H_J + H_{PJ}$, $H_J$ being the current-carrying helicity and $H_{PJ}$ the volume-threading one defined as:
\begin{gather}
    H_J = \int_V (\boldsymbol{A}-\boldsymbol{A_0})\cdot(\boldsymbol{B}-\boldsymbol{B_0})\hspace{0.2cm}dV\\
    H_{PJ} = 2\int_V \boldsymbol{A_0}\cdot(\boldsymbol{B}-\boldsymbol{B_0})\hspace{0.2cm}dV
\end{gather}
with $H_J$ measuring the twist, writhe and interwinding of the current-carrying field and $H_{PJ}$ the volume-threading helicity (between current-carrying and potential fields). More precisely, these separated components are individually gauge invariant but not conserved (only the total helicity is). This is because their dynamics is largely dominated by a transfer term. \citet{linan_time_2018} showed that in the course of an eruptive (CME-associated) reconnection scenario, $H_J$ tracks the transformation of the current-carrying field (i.e., in the form of a decrease) while $H_{PJ}$ (and thus $H_V$) might increase as a result of the partial transfer of $H_J$. This possible transfer is not assessed by our NLFF model, as the force-free assumption is not valid during magnetic reconnection. Also, a flare can leave behind a significant amount of current-carrying field (e.g, in partial eruptions), leading us to use pre- and post-impulsive phase one-hour averages to calculate a mean flare-related change of $H_J$. For completeness, we tested the impact of different lateral and vertical extents of the NLFF model volume on the helicity estimates, where we found minor variations. Following all of these aspects, we regard the change in $H_J$ as a reliable lower limit for the helicity ejected via the flare-associated CME. Limitations to this view are discussed in Section \ref{s:SumDis}.
\subsection{In-situ analysis}\label{ss:dataInSitu}
\subsubsection{Magnetic field measurements}\label{sss:InSituData}
At the time of the event, several spacecraft with magnetometers onboard were operating at different heliocentric distances: Solar Orbiter \citep[SolO/MAG;][]{muller_solar_2020, horbury_solar_2020} at $0.39$ AU, BepiColombo \citep[Bepi/MPO/MAG;][]{glassmeier_fluxgate_2010} at $0.58$ AU, STEREO-A (STEREO-A/MAG; \citet{acuna_stereoimpact_2008}) at $0.96$ AU, and Wind (Wind/MFI; \citet{lepping_wind_1995}) at $0.99$ AU. These magnetometers provide time series of the magnetic field direction and magnitude. In this study, we use the Radial-Tangential-Normal (RTN) coordinate system where the radial vector represents the axis from the Sun's center directed outward, the tangential vector is the cross product of the Sun's rotation axis with the radial vector, and the normal vector is the cross product of the other two vectors (i.e., $\boldsymbol{N} = \boldsymbol{R}\times \boldsymbol{T}$). The data for every spacecraft in the time frame between 2024 March 23 00:00 UT and 2024 March 27 12:00 UT are shown in Figure \ref{fig:allInSitu}.\\
\indent The ICME event analyzed in this study was detected by four spacecraft, as identified in the HELIO4CAST ICMECAT version 3.0 (see Footnote 1 p. 2, \citet{mostl_magnetic_2026} and \citet{figshare}) and in the DONKI catalog (ID 2024-03-23T01:25:00-CME-001). This ICME is then linked to two IPS arrivals at STEREO-A (2024-03-24T14:27:00-IPS-001) and Earth (2024-03-24T14:10:00-IPS-001), providing initial constraints on the structure and propagation of the associated magnetic flux rope (MFR). Among these spacecraft, BepiColombo was located at a substantial longitudinal separation from the Sun–Earth line (approximately $55^{\circ}$). The remaining three spacecraft were located at more moderate longitudinal separations, with Solar Orbiter (SolO) approximately $20^{\circ}$ from the Sun–Earth line, STEREO-A approximately $10^{\circ}$, and Wind near L1. Given this configuration, the spacecraft constellation would likely be capable of sampling a moderately inclined flux rope. In contrast, a highly inclined MFR would be less likely to produce signatures consistent with the observations across all spacecraft.
\begin{figure}[!ht]
\centering
\includegraphics[width=0.9\hsize]{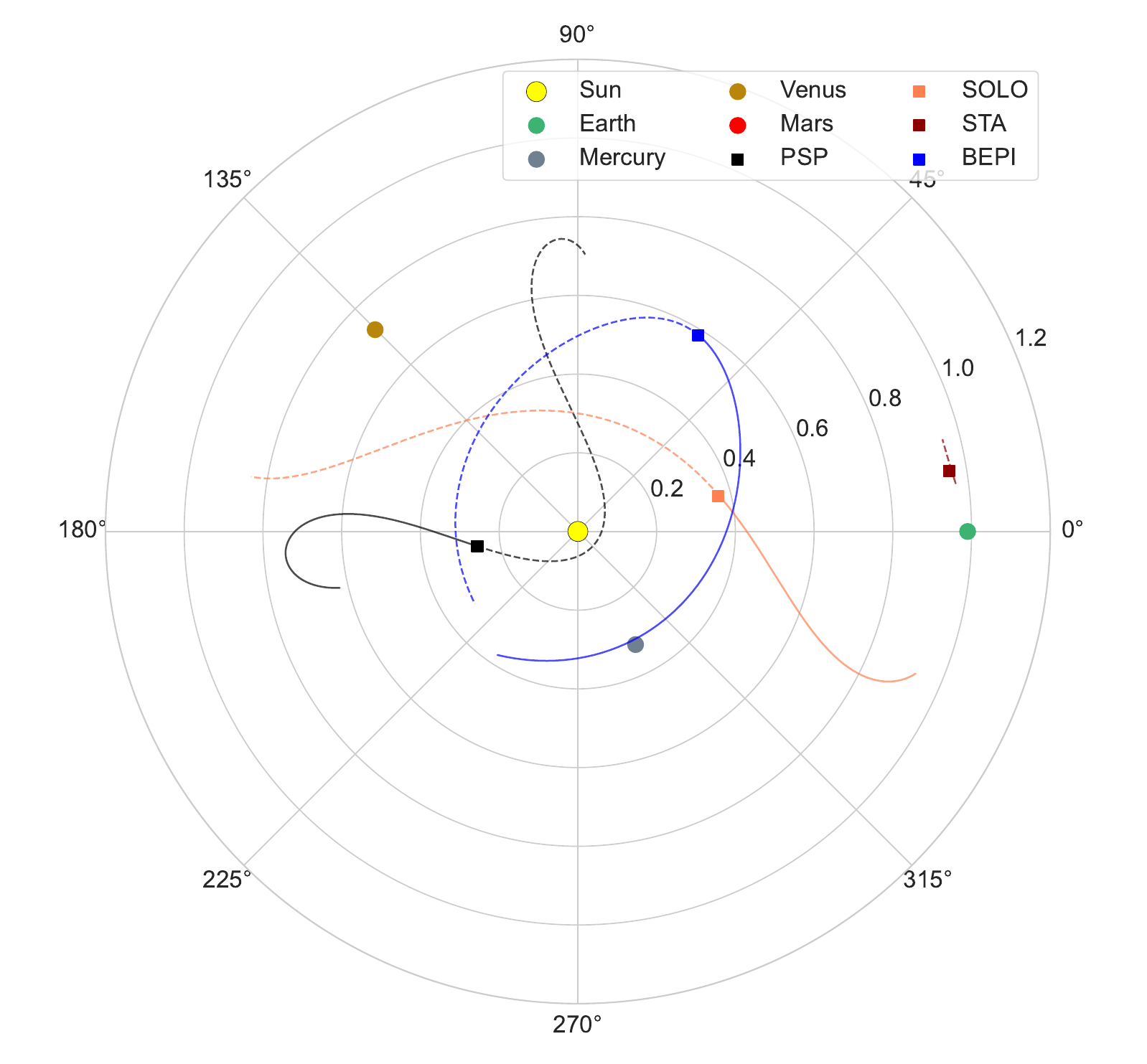}
    \caption{Spacecraft constellation when Solar Orbiter detected the magnetic cloud (2024-03-23 15:05~UT). Solid (dashed) lines represent the spacecraft trajectory 80 days before (after).}
        \label{fig:positions}
\end{figure}

\indent Using in-situ data, we are able to reconstruct the magnetic flux rope and retrieve key parameters which allow us to infer the magnetic helicity. In particular, the measured in-situ magnetic field data is fitted twofold: once based on the assumption of an underlying axially symmetric field corresponding to a linear force-free configuration (Lundquist), and once based on the assumption of an underlying uniformly twisted non-linear force-free magnetic field configuration (Gold-Hoyle in 3DCORE), as explained in detail in the following.
\subsubsection{Linear force-free field}
\begin{figure*}
    \sidecaption
    \includegraphics[width=0.35\linewidth]{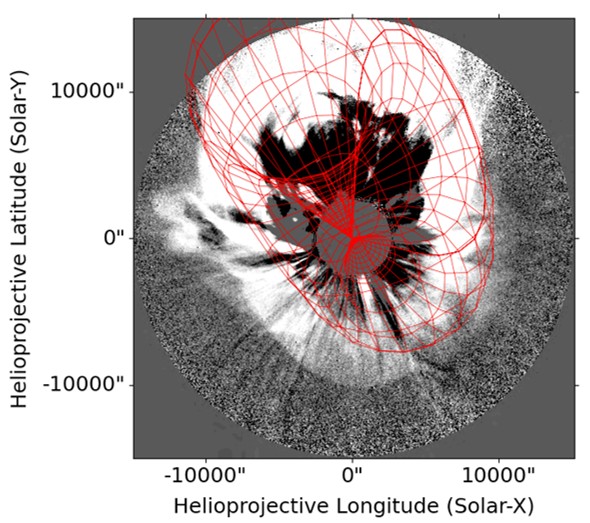}
    \includegraphics[width=0.34\linewidth]{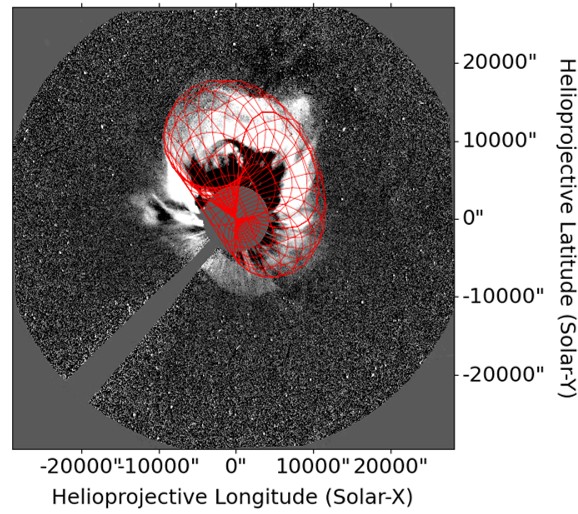}
    \caption{Graduated Cylindrical Shell (GCS) reconstruction of the 2024 March 23 CME. The background coronograph images were taken by STEREO-A/COR2 at 2024-03-23 03:23:30 UT (left panel) and by LASCO-C3 on 2024-03-23 at 03:30:07 UT (right panel).}
    \label{fig:GCS}
\end{figure*}
The Lundquist model \citep{lundquist1950} provides a solution to the force-free magnetic equation in cylindrical geometry and is often applicable to clear symmetrical ICME signatures \citep{demoulin_re-analysis_2019}. Here we use the expansion-modified force-free model, which is based on the Lundquist type of magnetic field, but adapted for the expansion \citep{palmerio_coronal_2024}. The Lundquist fitting is provided as one of the analysis modules of the Forbush decrease analysis tool\footnote{https://github.com/spearhead-he/FDAT} \citep[FDAT;][]{chikunova25}, which is a Python PyQt5-based interface designed for the analysis of ICMEs and their associated Forbush decreases. The model is fitted to the in-situ data, and the fitting procedure yields several key flux rope parameters, including the axis orientation, the axial magnetic field strength, the chirality (or handedness) of the flux rope, and the impact parameter. Additional quantities, such as the initial and final flux rope radius, expansion speed, and the axial and poloidal magnetic fluxes at the point of measurement, can also be derived. To perform the fitting, initial guesses are required for the flux rope handedness ($\sigma_H$, which determines the sign of the magnetic helicity), the axis direction in latitude ($\theta_0$) and longitude ($\phi_0$), the impact parameter ($p_0$), the central magnetic field strength ($B_0$), and the flux rope expansion time ($T_0$). The initial values adopted in this study are computed using a Graduated Cylindrical Shell \citep[GCS;][]{thernisien_modeling_2006} reconstruction on the CME observed by the LASCO-C3 \citep{brueckner_large_1995} and STEREO/COR2-A \citep{kaiser_stereo_2008} white-light coronographs (a representation of the latter is shown in Figure \ref{fig:GCS}). The initial guesses used for the Lundquist fitting are listed in Table~\ref{tLundquist}. This helicity computation is based on the theoretical work of \citet{Devore_2000} and more specifically on its equation (49):
\begin{equation}\label{eq:lundquist}
    H_L = 0.60\cdot\sigma_H\cdot B_0^2 \cdot R^3 \cdot l
\end{equation}
where $R$ and $l$ are the radius and length, respectively, of the flux rope obtained with the Lundquist fit, and $H_L$ is the helicity inferred with the Lundquist model. Finally, to give a comparison with other methods and assess the reliability of the result, we calculate an error based on the propagation of the instrument-considered precision ($0.05$ doy for the time, $10~\text{km s}^{-1}$ in velocity and $1~\text{nT}$ in magnetic field intensity).
\begin{table}[ht]
\caption{\label{tLundquist}Initial parameter estimates for the Lundquist fit.}
\centering
\begin{tabular}{lccccl}
\hline\hline
Parameters&SolO&BEPI&STEREO-A&Wind&Units\\
\hline
$\theta_0$    &$-80$ &$-80$ &$-60$ &$-60$   &deg\\
$\phi_0$     &$5$   & $5$ &$5$  &$5$   &deg\\
$p_0$  &$-0.3$   & $-0.3$ &$-0.3$ &$-0.3$   & --\\
$B_0$ & $150$&$80$ &$30$& $30$& ${\rm nT}$\\
$T_0$ &$15$ & $20$&$40$ &$40$ & ${\rm h}$\\
$R_{S}$& $0.39$ & $0.58$ &$0.96$ & $0.99$ & AU\\
\hline
\end{tabular}
\tablefoot{Initial parameter estimates used for the Lundquist fitting with the direction in latitude ($\theta_0$), the longitude ($\phi_0$), the impact parameter ($p_0$), the central magnetic field strength ($B_0$) and finally the duration of the MFR crossing ($T_0$). The last row shows the radial distance from the Sun ($R_S$). The handedness is always considered negative (left-handed), as this type of flux rope matches the typical left-handed in-situ profiles and emerges from a left-handed flux rope at the solar active region.}
\end{table}
\subsubsection{Uniformly twisted field}
\indent Following the results of \citet[chap. ~3.2]{dasso_new_2006}, and specifically Eq.(11):
\begin{equation}\label{eq:GH}
    H_{GH} = \frac{8\pi[ln(1+U^2/4)]^2}{U^4} \cdot B_0^2 \cdot R^4 \cdot \tau_0
\end{equation}
where $U=2\cdot \tau_0 \cdot R$, it is possible to infer helicity directly from a Gold-Hoyle-like flux rope model \citep{Gold_Hoyle_1960} providing the flux tube radius $R$, magnetic field intensity at the center $B_0$, twist per unit length $\tau_0$ and length of the MFR $l$, the latter making the link between the twist factor from 3DCORE and the twist per unit length. These parameters are retrieved using the semi-empirical model 3DCORE, which relies on the assumption of a uniformly twisted non-linear force-free magnetic field. This model was first introduced in \cite{mostl_forward_2018} and it evolved later \citep{weiss_analysis_2021, weiss_multi-point_2021} to the version we are using in this study, described in \citet{rudisser_understanding_2024, davies_flux_2024, rudisser_towards_2026}.\\
\indent 3DCORE assumes a torus-like geometry with an embedded Gold-Hoyle-like magnetic field configuration \citep{Gold_Hoyle_1960}. This flux rope has an elliptical cross-section that expands self-similarly while having both ends attached to the Sun. The propagation is described by a drag-based model \citep{vrsnak_propagation_2013} applied to the frontal part of the CME, with a constant solar wind speed being one of the model's input parameters. This model infers the flux rope parameters using an Approximate Bayesian Computation-Sequential Monte-Carlo (ABC-SMC) method \citep{weiss_analysis_2021}. The input parameter ranges can be constrained by physical considerations or measurements and are used as priors for the ABC-SMC fitting procedure. The algorithm then samples possible 3DCORE parameter combinations and iteratively narrows them down to best reproduce the observed in-situ magnetic field profiles. Table \ref{t1} shows the input parameter ranges we chose for this CME at all spacecraft. The ranges are kept broad on purpose except for the latitude and longitude that are defined directly from the DONKI event catalog (ID 2024-03-23T01:25:00-CME-001), with a $\pm40^{\circ}$ ($\pm30^\circ$) range in longitude (latitude).
\begin{table}[ht]
\caption{\label{t1}3DCORE parameter ranges.}
\centering
\begin{tabular}{lccl}
\hline\hline
Parameters&Minimum&Maximum&Units\\
\hline
Longitude    &$-18$       &$62$   &deg\\
Latitude     &$-28$       &$32$   &deg\\
Inclination  &$0.0$       &$360.0$   &deg\\
$d_{1au}$    &$0.05$      &$0.35$    &AU\\
$\delta$     &$1.0$       &$3.5$     &\\
$T_{fac}$    &$-250$      &$250$     &\\
$v_0$        &$1513$      &$1713$   &${\rm km\,s}^{-1}$\\
$B_{1au}$    &$5.0$       &$65.0$    &nT\\
$\Gamma$     &$0.2$       &$3.0$     &\\
$v_{sw}$     &$405$       &$605$   &${\rm km\,s}^{-1}$\\
$n_a$        &$1.0$       &$2.0$     &\\
$n_b$        &$1.0$       &$2.0$     &\\
\hline
\end{tabular}
\tablefoot{The listed input parameters are: longitude and latitude of the flux rope (given in HEEQ coordinates), the inclination (an inclination of $0^\circ[180^\circ]$ would mean an MFR propagating central axis parallel to the ecliptic), the diameter of the cross section at 1 AU ($d_{1au}$), the aspect ratio of the cross section at 1 AU ($\delta$), the twist factor ($T_{fac}$) which is related to the number of twists over the whole torus, the launch speed ($v_0$), the magnetic field strength at 1 AU ($B_{1au}$), the drag parameter ($\Gamma$), the global solar wind speed ($v_{sw}$), the expansion rate ($n_a$) of the cross-section diameter, and the magnetic field decrease rate ($n_b$).}
\end{table}\\
\indent To fit 3DCORE to these measurements, we define a set of fitting points during the MFR interval. At these points, the modeled magnetic field is fitted to the observed in-situ data, and the difference between them is quantified using a normalized root mean square error (N-RMSE). This N-RMSE is then minimized during the fitting process to find parameter combinations that best reproduce the measured magnetic field profile. We use the same number of fitting points (i.e., $7$) in every set placed empirically, with the objective of reflecting the rotating trend of the magnetic field in the fit. We chose this setup because the number of fitting points does not alter the result, and to avoid fitting perturbations or non-rotating parts \citep[for more details, see][]{weiss_multi-point_2021}. 3DCORE provides the number of twists in the reconstructed torus-like MFR. We then consider a classical representation of a MFR structure where all the twist is contained in the front region linked to the Sun by two untwisted legs. Following \citet{Devore_2000}, we consider a flux rope with an angular width (angular size of the front region) between $25^\circ$ and $35^\circ$. Also, the MFR radius changes depending on the position on the torus. For this reason, we choose to compute helicity by averaging the results of the helicity computation with the flux rope's apex radius and with the radius on the side of the twisted front region (e.g. the radius is taken at $0^\circ$ and $15^\circ$ for a $30^\circ$ front region). This allows us to account for the impact of the geometry on the helicity results obtained (see section \ref{ss:MFRevol}) and to compute an error estimate of the newly inferred helicity.
\section{Results}
\subsection{Solar source and eruptive activity}\label{ss:eruption}
\begin{figure*}[!ht]
\centering
\includegraphics[width=0.8\hsize]{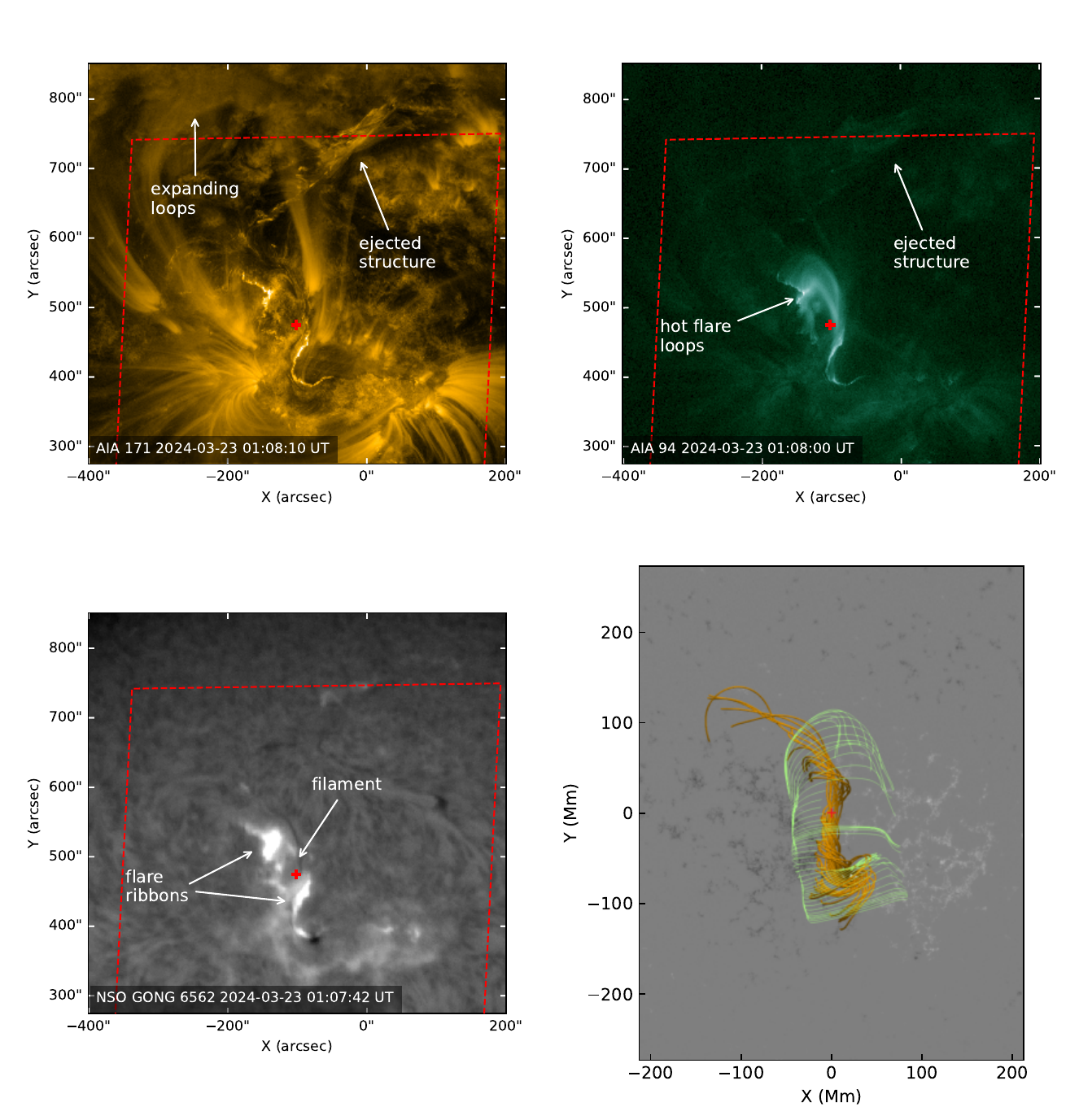}
    \caption{NOAA AR13614 during the impulsive phase of the SOL-2024-03-23T00:58X1.1 flare. Panels a, b and c show AIA~171~$\AA$, AIA~94~$\AA$ and NSO GONG H-$\alpha$ images, respectively. Coordinates are in arcseconds with respect to Sun center. Red contours and crosses portray the limit and center of the vector magnetogram used for NLFF extrapolation. Panel d displays one reconstruction at 2024-03-23 01:12 UT, with its center also marked by a red cross for readability. The field line sampling in this last panel shows the connectivity of the main positive region (orange lines) along with the surrounding field (green lines).}\label{fig:corona}
\end{figure*}
\begin{figure}[ht]
    \centering
    \includegraphics[width=0.85\linewidth]{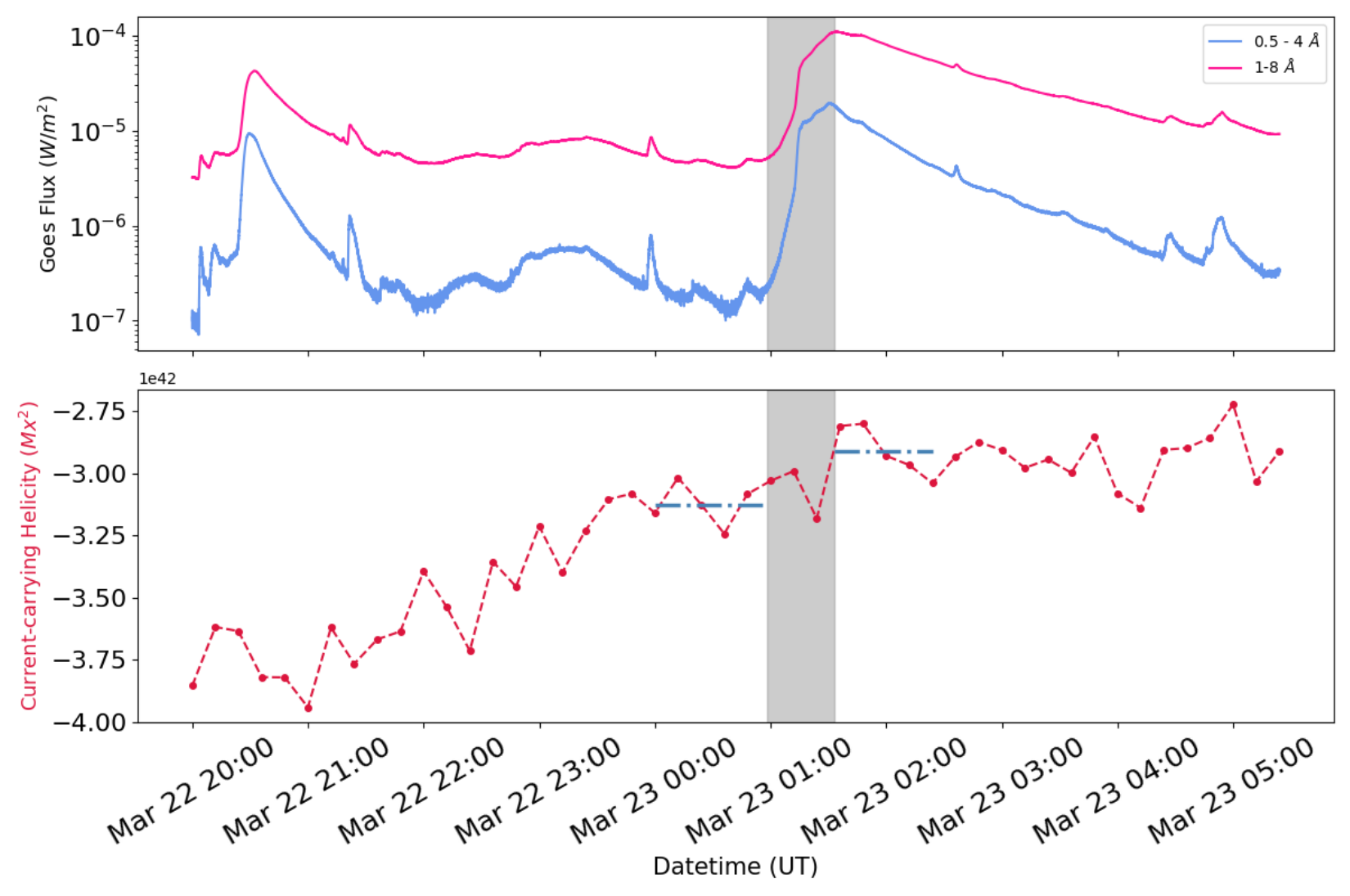}
    \caption{X-ray and magnetic helicity evolution of the 2024 March 23 flare. Top: GOES 18 soft X-ray flux ($0.5\text{ - }4~\AA$ in blue, $1\text{ - }8~\AA$ in pink). Bottom: current-carrying magnetic helicity. The time between the start of the flare (2024 March 23 00:58 UT) and its peak (2024 March 23 01:33 UT) is shown as a gray shaded area. The helicity mean pre-flare and post-flare values are shown as blue dash-dotted lines over the timesteps taken in their respective calculation.}\label{fig:HBudget}
\end{figure}

Figure \ref{fig:corona} shows chromospheric and coronal emission at the beginning of the impulsive phase, along with a field line sampling of a pre-flare NLFF extrapolation result. In detail, panels a and b show, respectively, the EUV emission at 171 \AA and 94\AA of NOAA~AR13416 during the early impulsive phase of the X1.1 flare that started at March 23 00:58~UT and peaked at March 23 01:33~UT. In panel c, we display an H-$\alpha$ image at a similar time. The last panel (panel d) shows a result of the NLFF model at a similar time with field lines highlighted.\\
\indent During the flare, a pre-existing filament partially erupted. In H-$\alpha$ observations (panel c), distinct flare kernels/ribbons are observed in the form of bright patches north of the sunspot during the early impulsive phase, on both sides of the erupting filament. Near-in-time AIA~171~\AA~observations (panel a) exhibit thin rims of bright emission co-spatial with the H-$\alpha$ ribbons. The bright overarching coronal loops are observed to expand with time. Both H-$\alpha$ and coronal observations show that an apparently twisted structure is ejected in a north-western direction. In addition, AIA~94~\AA~observations (panel b) show hot cusp-shaped flare loops connecting the bright flare ribbons. Prior to the flare, a dark filament extends from near the central sunspot towards the north, approximately 200~arcseconds in length. Part of the filament is still (again) visible after the flare, shorter in length, with its extensions toward the sunspot and the quiet-Sun region in the north apparently removed. The pre-flare coronal structure presents an inverse S-shaped sigmoid and reverse-J-shaped flare ribbons supporting a left-handed MFR.\\
\indent Co-spatial with the observed filament structure, we recover a magnetic flux rope in our pre-flare NLFF models (see Figure \ref{fig:corona}d for a snapshot at 01:12 UT). Orange-colored sample field lines outline the connectivity of the main positive polarity patch in the underlying photospheric vector magnetogram (grayscale background), while those colored in green represent the surrounding field. The NLFF model suggests a left-handed (i.e., negative twist) flux rope structure supporting the filament material observed in the EUV and H-$\alpha$ images.\\
\indent We used a qualifying NLFF model to estimate the current-carrying helicity in a $\sim$9-hour time window centered around the flare start time (see Figure \ref{fig:HBudget}). These models show a satisfying force-free and solenoidal quality (i.e., values of $\theta_J<21^\circ$ and $E_{div}/E<0.07$, respectively). Using the current-carrying helicity values prior to and after the impulsive phase of the flare, we compute one-hour averages resulting in $H_J=-3.12\pm 0.08 \times 10^{42}~\text{Mx}^2$ before the flare start and $H_J=-2.91\pm 0.10 \times 10^{42}~\text{Mx}^2$ after the flare peak. Consequently, we find a flare-associated mean change in the current-carrying helicity of $\Delta H_J = -2.16\pm1.32\times10^{41}~\text{Mx}^2$, a value in line with previous studies on eruptive X-class flares based on similar approaches and methods \citep[see][]{thalmann_tracking_2023}.
\subsection{Upper corona and interplanetary space}\label{ss:MFRevol}
\begin{figure*}[!htp]
    \centering
    \includegraphics[width=0.85\linewidth]{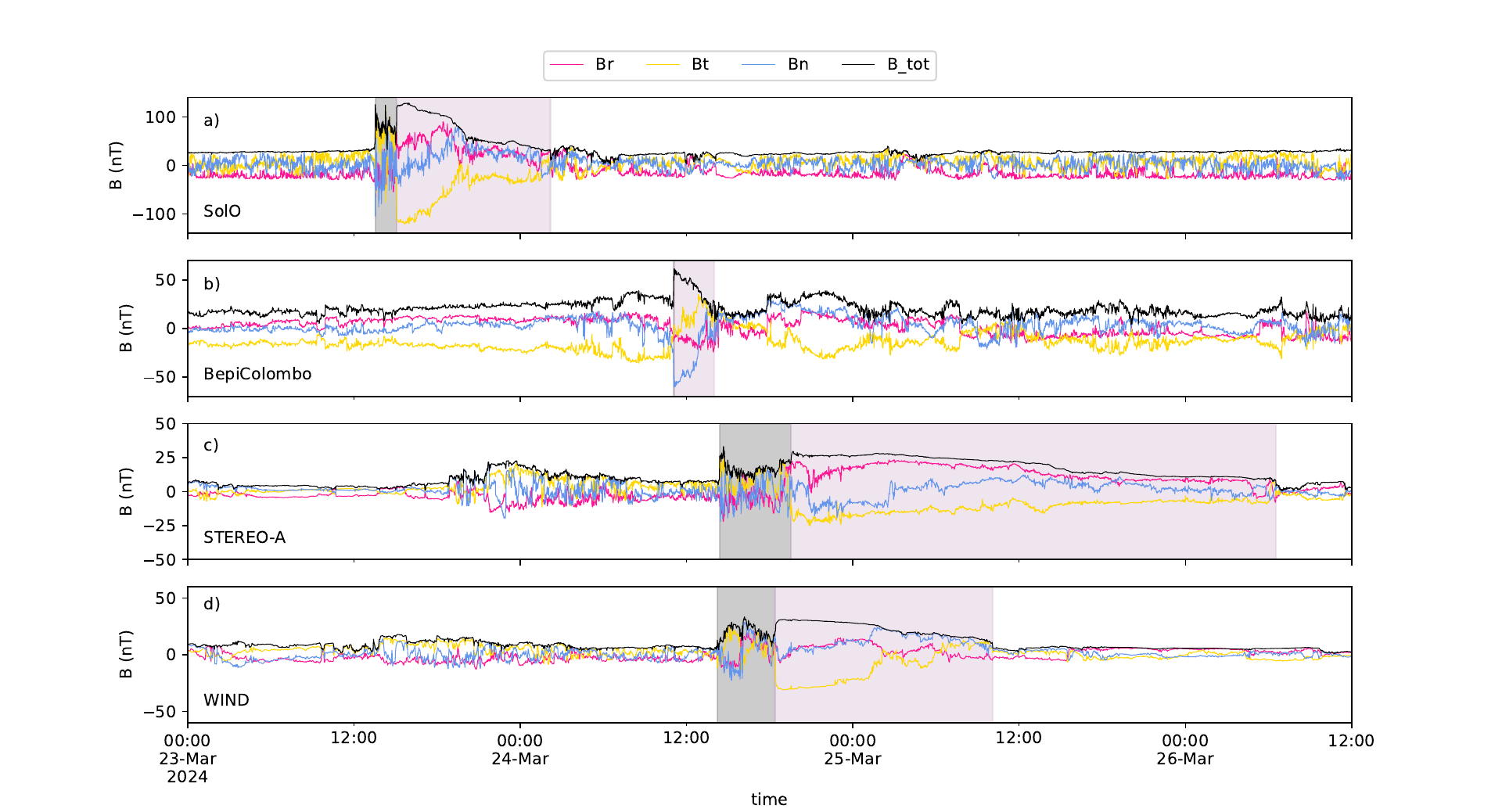}
    \caption{Time series (between 2024 March 23 00:00 UT and 2024 March 26 12:00 UT) of the ICME magnetic field components in RTN coordinates for the spacecraft observing the event (Solar Orbiter, BepiColombo, STEREO-A and Wind, from top to bottom). The radial component (Br; pink) is aligned with the Sun-spacecraft line, and the tangential direction (Bt; yellow) is defined through the projection of the solar rotational axis onto the plane perpendicular to the radial component. The normal direction (Bn; blue) points along the projection of the solar north pole. The total magnetic field magnitude is shown in black. The gray shaded region highlights the CME magnetic sheath, while the purple region highlights the magnetic flux rope.}\label{fig:allInSitu}
\end{figure*}
\begin{figure*}[!ht]
    \centering
    \includegraphics[width=0.25\linewidth]{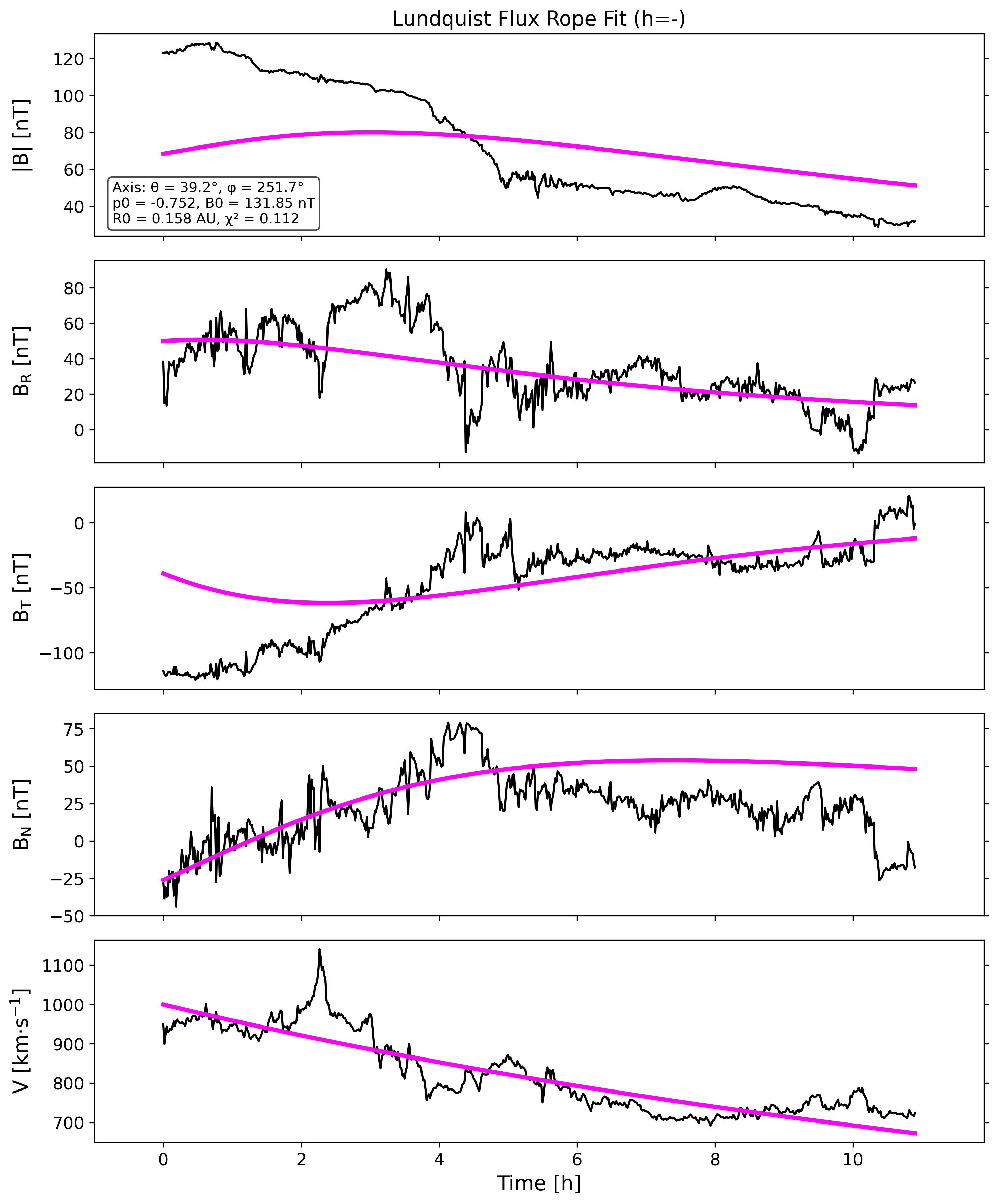}
    \includegraphics[width=0.26\linewidth]{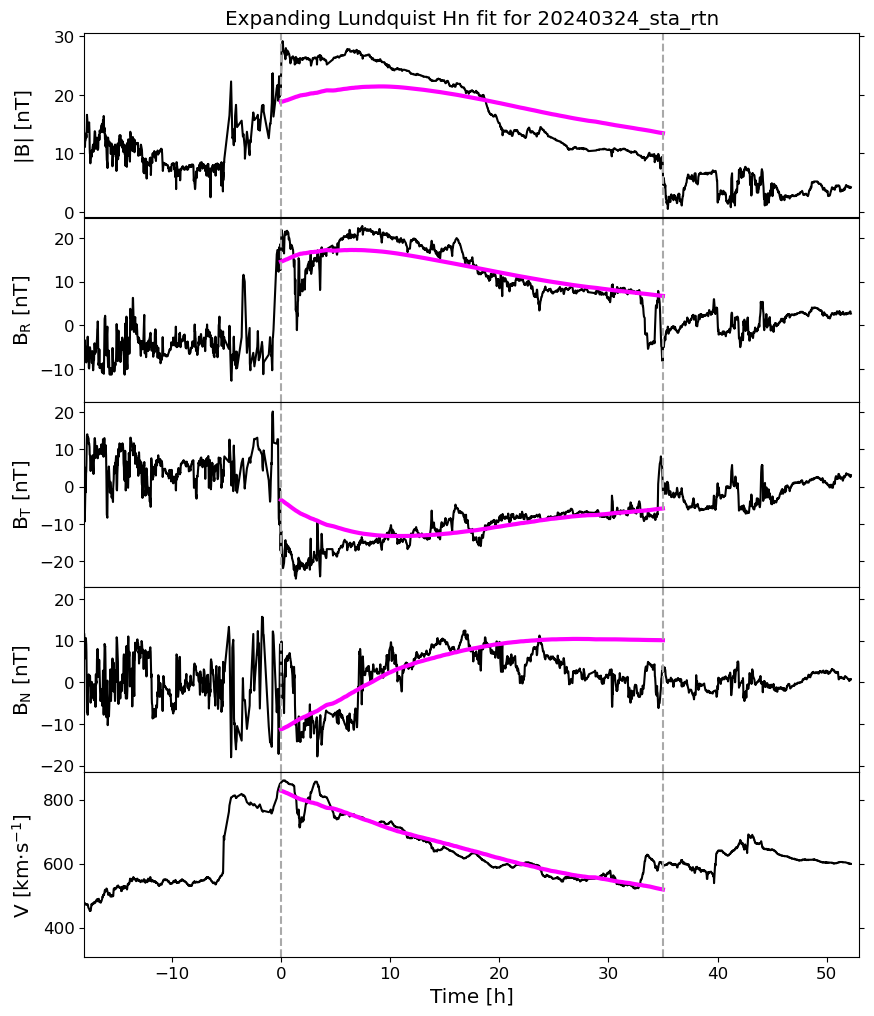}
    \includegraphics[width=0.25\linewidth]{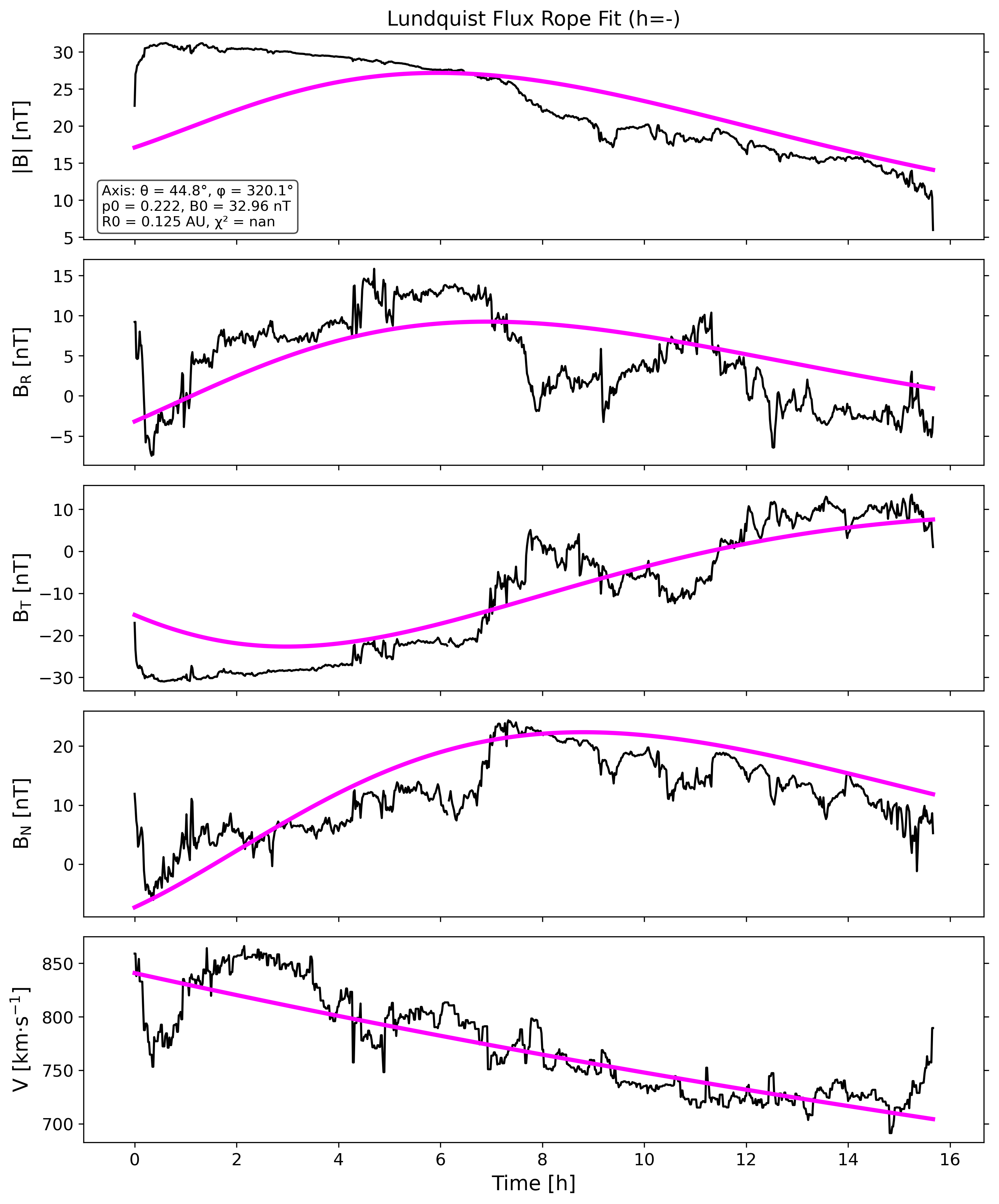}
    \caption{Lundquist fit curves (pink) overlaying the in-situ magnetic field data (black) for Solar Orbiter (left), STEREO-A (middle), and Wind (right). For every reconstruction, the top panel shows the result for the magnetic field magnitude. The three following show the RTN components in order, and the bottom one shows the speed profile of the event. The reference time marked as time 0 is the start of the MFR (purple shaded area in Figure \ref{fig:allInSitu})}\label{fig:allLUND}
\end{figure*}
\begin{figure}[!ht]
    \centering
    \includegraphics[width=0.8\linewidth]{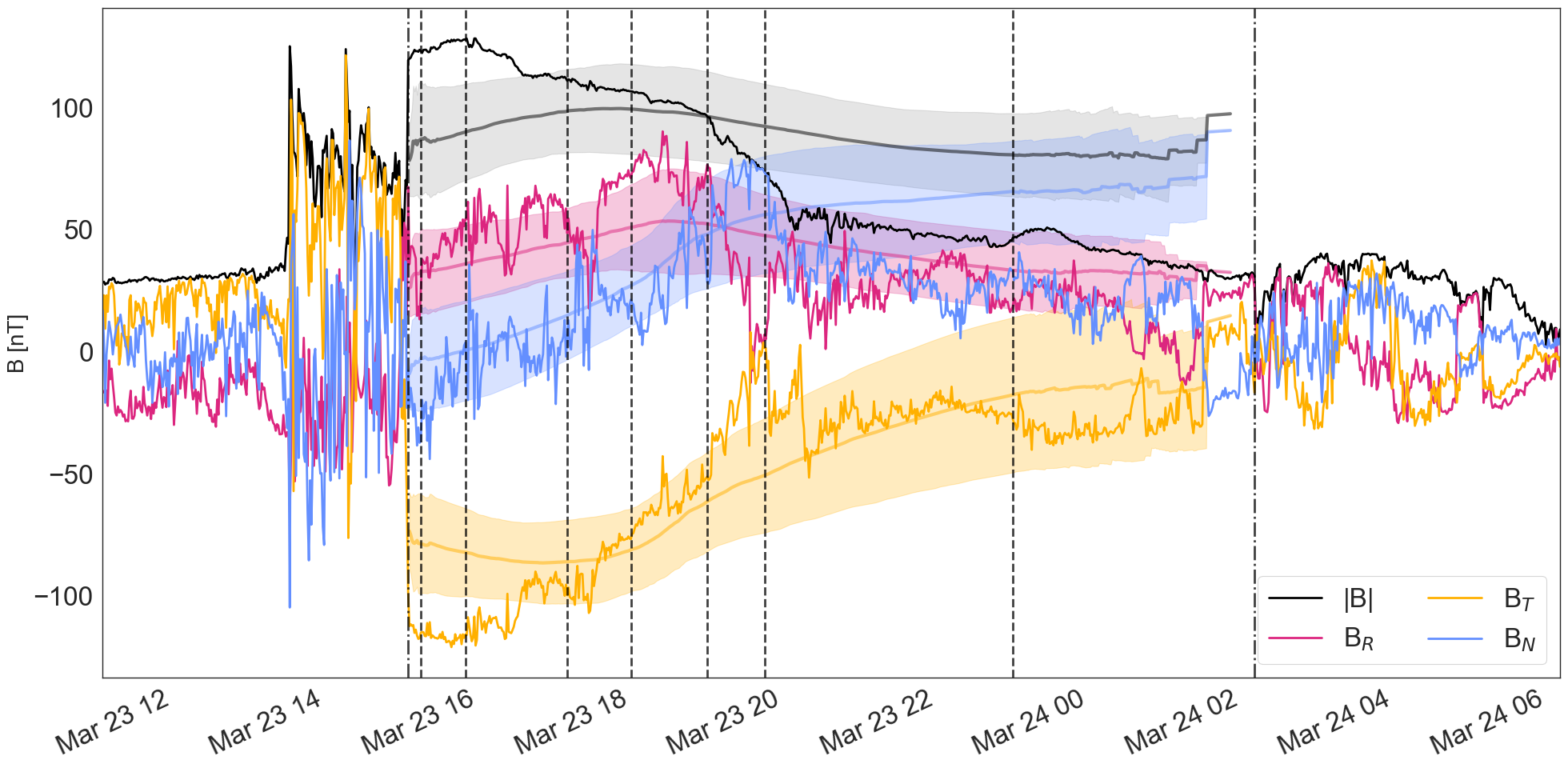}
    \includegraphics[width=0.8\linewidth]{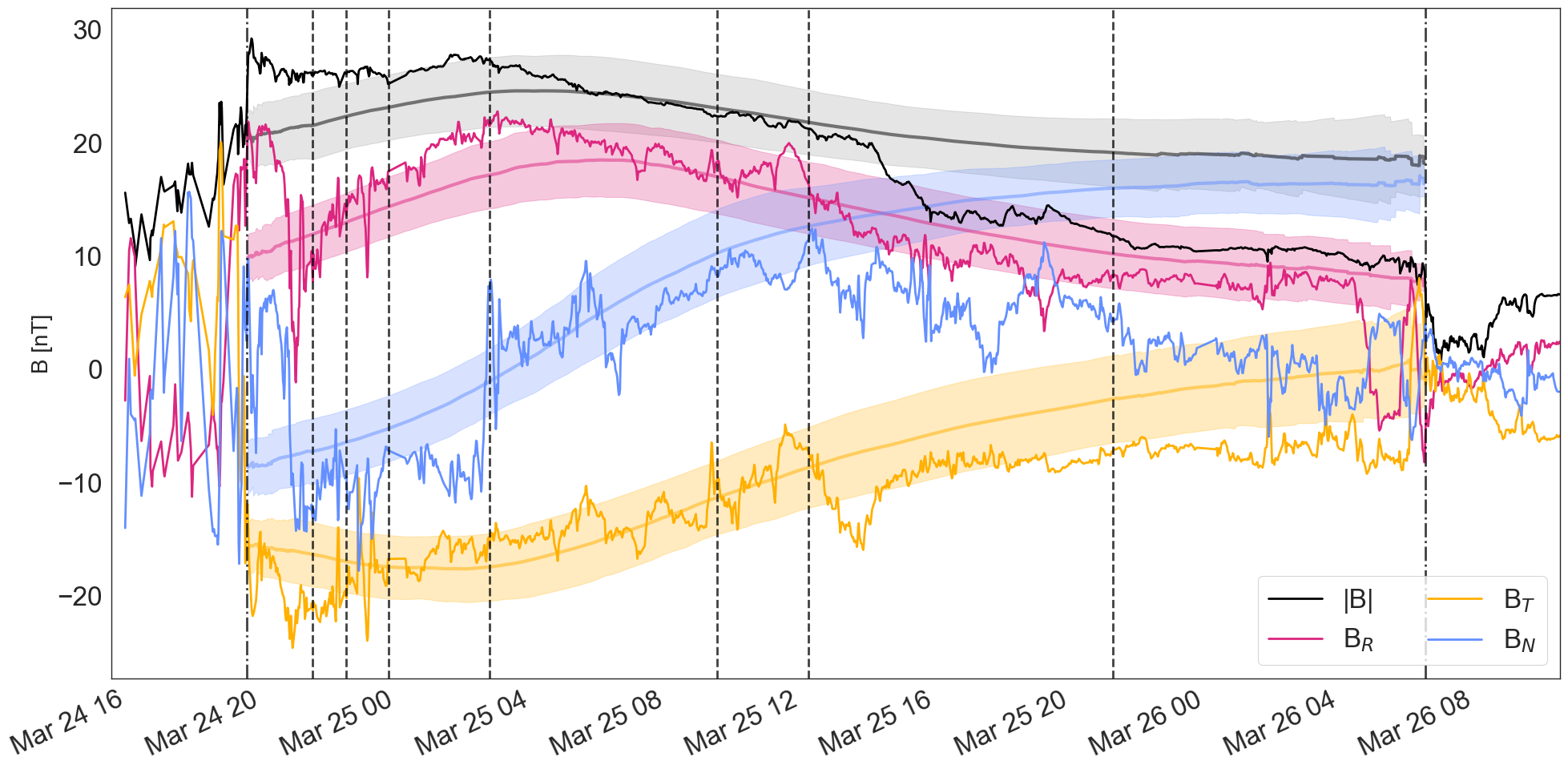}
    \includegraphics[width=0.8\linewidth]{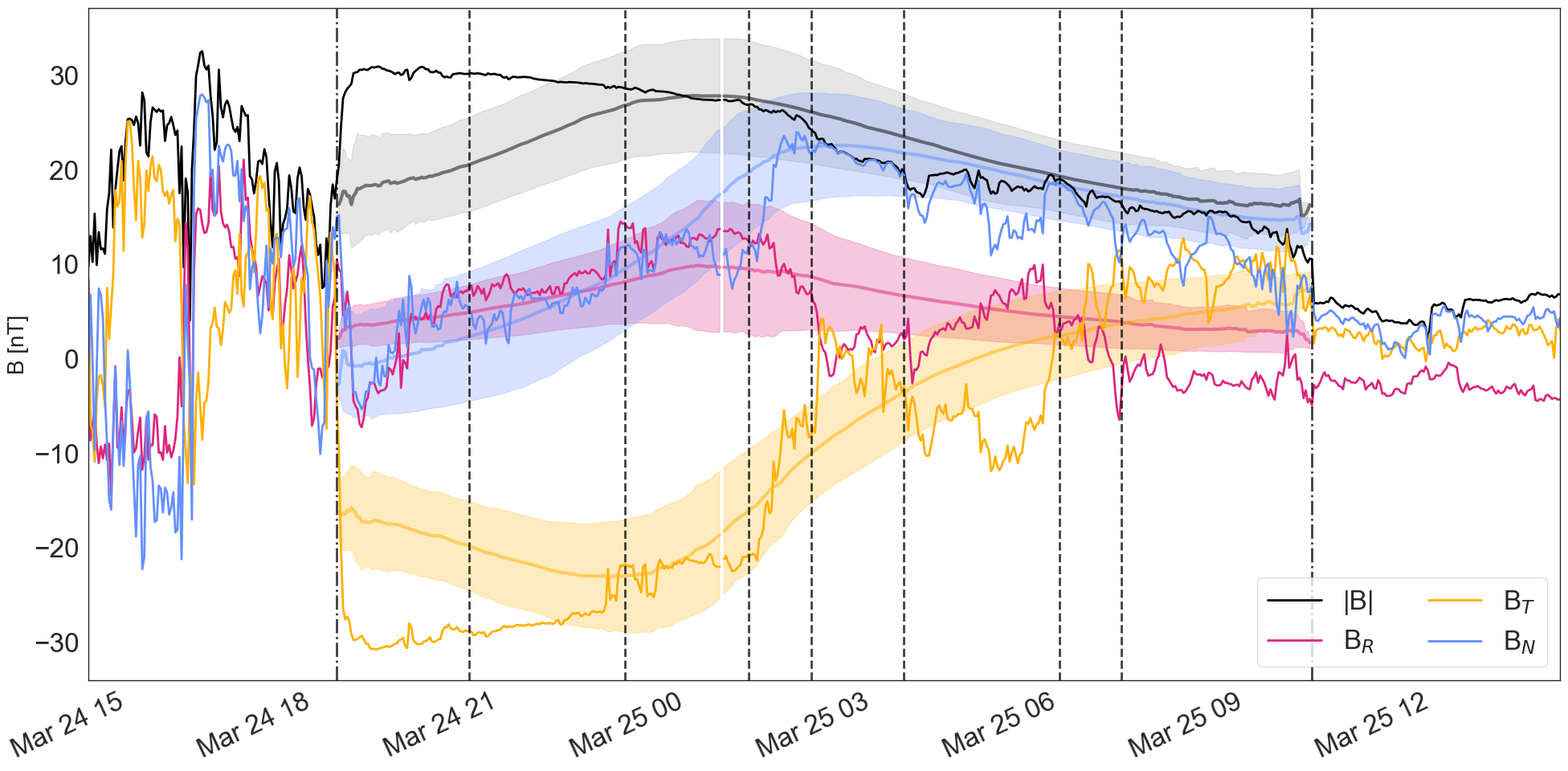}
    \caption{Results of the 3DCORE fitting for each spacecraft (Solar Orbiter in the top panel, STEREO-A in the middle one and Wind in the third panel) compared to the magnetic field measurement in RTN coordinates (as in Figure \ref{fig:allInSitu}). The colored shaded areas are the 2$\sigma$ spread of the ensemble for every component (including the magnitude, in black), meanwhile the solid line represents the mean of the ensemble. The dash-dotted lines delimit the MC signature, whereas the dotted ones represent the chosen fitting points.}\label{fig:allfit}
\end{figure}

The March 23, 2024 ICME event is captured on its way through the heliosphere by Solar Orbiter (at 0.39 AU from the Sun), BepiColombo (at 0.58 AU), STEREO-A (at 0.96 AU), and Wind (at 0.99 AU) with a small angular separation between Solar Orbiter and the near-L1 spacecraft (see Figure \ref{fig:positions}). Figure \ref{fig:allInSitu} shows the in-situ magnetic field components of the ICME measured by different spacecraft. They clearly reveal the shock front, characterized by a jump in magnetic field magnitude followed by characteristic turbulent sheath properties (gray shaded area), and MC signatures with a rotation of the magnetic field (bipolar blue Bn component here) and a steady decrease in the magnetic field total intensity (B\_tot in the purple shaded area). The in-situ data of Solar Orbiter, STEREO-A, and Wind show great similarity (panels a, c and d in Figure \ref{fig:allInSitu}, respectively), as their sheath region is clearly identified and the components show comparable trends. Though not explicitly shown, we note that the shock arrival is followed by a period of increased density, temperature, and plasma beta (these data are not available for BepiColombo). Together with a highly fluctuating magnetic field, these observations are indicative of a sheath region before the arrival of the MFR \citep{zurbuchen_-situ_2006, kilpua_coronal_2017}. In the MC region (purple-shaded area), the bipolar Bn component is visible, along with an evolution of the Bt component from negative to near-zero values. We can also notice the general expansion of the MC region between near-Sun spacecraft (SolO) and near-L1 ones (STEREO-A, Wind). On the other hand, BepiColombo (panel b) displays very different characteristics, with no turbulent sheath but only a shock-like jump in magnetic field intensity, indicating that it most likely observed the inner/core ICME structure from its side. The shock arrival is observed at Solar Orbiter around 23 March 13:00 UT, at BepiColombo around 24 March 11:04 UT and at STEREO-A and Wind around 24 March 13:00 UT. \\
\indent Looking only at the magnetic field rotation in Solar Orbiter's measurements (top panel), it is clear that the Bt (yellow line) component exhibits a clear half-rotation. Bn also exhibits a slight rotational pattern from negative to positive values. This is in line with the definition of an ENW type of MFR \citep{bothmer_structure_1998}, i.e., left-handed and intermediate to highly inclined. These observations are in line with the pre-flare coronal structure described in Section \ref{ss:eruption}. Regarding the intermediate-to-high inclination signature, GCS reconstruction along with coronographic observations of the CME by LASCO white-light and by STEREO-A/COR2 support this guess with a tilt of roughly $60^\circ$. STEREO-A and Wind measurements (3rd and 4th panel from the top, respectively) are in line with these observations, with a similar Bn (blue line) rotation and Bt (yellow line) half-rotation. The main differences between signatures observed by STEREO-A and Wind are a different expansion, where the duration of the MC at STEREO-A is longer than that observed at Wind (see yellow area in panels c and d in Figure \ref{fig:allInSitu}), and a difference in the strength of the radial component (Br in the same panels), where STEREO-A measures higher values.\\
\indent In the BepiColombo in-situ data (panel b), different features are observed. First, there is only a half-rotation observed in Bn, moving from negative values to zero. Second, the rotation in Bt (yellow line) is observed at mostly positive values. Third, the radial component (pink line) exhibits a transition from positive to negative values. These observations and the longitudinal position of BepiColombo indicate that it most likely probed the leg of the ICME \citep[see][]{owens_legs_2016}. Due to the absence of a sheath region and the lack of plasma data, paired with these differences in the magnetic field rotational pattern, we do not further include the BepiColombo data in our analysis. For completeness, the analysis of the BepiColombo data is appended (Appendix \ref{ap:BEPIresults}).\\
\indent We first model this event with a Lundquist-type model based on a simple, circular cross-section containing a linear force-free magnetic field \citep{demoulin_re-analysis_2019}. The ICME expansion is assumed to be radially self-similar between Solar Orbiter and Wind, which is in line with the monotonically decreasing speed profiles (bottom panels in Figure \ref{fig:allLUND}). The Lundquist fit model input parameters are estimated by using images of the CME early stages along with the spacecraft measurements and are available in Table \ref{tLundquist}. To evaluate the goodness of the fit, we compare the $\chi^2$ metrics obtained for every fit (see Table \ref{t2}). We note that the fit quality is similar for STEREO-A and Wind, but the impact parameter is only acceptable for the latter. The difficulties of fitting the observations with this model might come from a high non-self-similar expansion where the MFR evolved away from the classical model assumed by this method, especially for spacecraft near 1 AU. Based on the limited performance of the Lundquist fit, we can understand the strong deviations of the resulting helicity values we compute, also shown in Table \ref{t2} with the label $H_L$ ($p_L$ for the impact parameters). The impact parameter, for STEREO-A and Solar Orbiter, is large ($\gtrsim0.7$), indicating that the spacecraft would probe the reconstructed MFR far away from its axis ($0$ being the crossing of the axis), which may also partially explain the strongly deviating findings. For Wind observations, we find a value for the magnetic helicity derived with the Lundquist model of $H_L=-5.01 \pm 0.8 \times 10^{41}$~\text{Mx}$^2$. The helicity values for STEREO-A and Solar Orbiter are substantially larger, by a factor of $\sim10$ and $\sim10^2$, respectively. The challenging task of fitting these observations and the assumptions behind this model encourage us to compare the results with a more sophisticated approach.\\
\indent It is possible to infer the helicity content of a MC with different assumptions. Hence, we also use a non-linear force-free uniformly twisted model, 3DCORE, as an alternative approximation for the MFR, in order to compute the helicity. Figure \ref{fig:allfit} presents the fitting results of 3DCORE for each magnetic field component at Solar Orbiter, STEREO-A, and Wind (for BepiColombo, see Appendix \ref{ap:BEPIresults}) as a shaded area representing the $2\sigma$ spread of the ensemble. The solid lines are the measured magnetic field, in the spacecraft RTN coordinates, same as displayed in Figure \ref{fig:allInSitu}. The solid lines inside the shaded area represent the mean of the ensemble used for helicity calculations. The event start and end are shown by two black vertical dashed-dotted lines, and the gray dashed lines are the chosen fitting points.\\
\indent The results obtained with 3DCORE (see Table \ref{t2}) show good consistency between SolO, STEREO-A and Wind within the uncertainty ranges, especially regarding the MFR location (i.e. a negative longitude, a near-equator latitude and a high inclination), but also in a vast majority of parameters ($B_{1au}$ around $25~\mathrm{nT}$ and $n_b$ around $1.6$). Two parameters, $v_0$ and $v_{sw}$, are strongly constrained with observational data and are thus also in good agreement for each spacecraft. This agreement shows the strength of this comparative study with multi-spacecraft observations of the event. In particular, the diameter at 1 AU between STEREO-A and Wind is substantially different ($0.28$ and $0.16$, respectively) even though they are at a comparable radial distance from the Sun (0.96 and 0.99 AU, respectively). The same difference can be seen in the aspect ratio of the cross-section ($\delta$), with STEREO-A being around $1.25$, while Solar Orbiter and Wind are around $2.3-2.4$. STEREO-A and Wind therefore show a different geometrical evolution of the ICME, with a much larger cross-section at STEREO-A, and with Wind being closer to the representation at Solar Orbiter. Here again we can compare the quality of the fit with the N-RMSE values (see Table \ref{t2}). They suggest an acceptable capacity of the model in fitting the observed data. The requirements we propose for the fitting process (7 user-specified positions to grasp important aspects in the in-situ profiles) are a possible explanation for why a higher fitting quality was not reached \citep[see][]{rudisser_towards_2026}.
\begin{table*}[ht]
\caption{\label{t2}Results and quality (N-RMSE) of 3DCORE reconstructions for all parameters (detailed in Table \ref{t1}).}
\centering
\begin{tabular}{lccccl}
\hline\hline
Parameters & {SolO} & {STEREO-A} & {Wind} & {BepiColombo} & units \\
\hline
Longitude    & $-13 \pm 4$ & $-12 \pm 4$ & $-11 \pm 3$ & $33 \pm 5$ & deg \\
Latitude     & $1 \pm 6$   & $-6 \pm 2$  & $3 \pm 5$   & $24 \pm 5$ & deg \\
Inclination  & $329 \pm 6$ & $341 \pm 4$ &$ 313 \pm 7$ & $131 \pm 6$ & deg \\
$d_{1au}$    & $0.33 \pm 0.01$   & $0.28 \pm 0.03$   & $0.16 \pm 0.01$   & $0.07 \pm 0.01$ & AU \\
$\delta$     & $2.36 \pm 0.63 $  & $1.25 \pm 0.18$   & $2.55 \pm 0.62$   & $2.79 \pm 0.5$ & \\
$T_{fac}$    & $-18.2 \pm 6.2$ & $-23.1 \pm 4.4 $& $-48.8 \pm 12.1$ & $165.9 \pm 44.4$ & \\
$v_0$        & $1620 \pm 53$     & $1625 \pm 54$     & $1618 \pm 52$     & $1614 \pm 56$ & {$\mathrm{km\,s^{-1}}$} \\
$B_{1au}$    & $20 \pm 5 $ & $25 \pm 2 $ & $28 \pm 3$   & $19 \pm 3$ & nT \\
$\Gamma$     & $0.43 \pm 0.12$   & $0.25 \pm 0.03$   & $0.27 \pm 0.04$   & $1.8 \pm 0.67$ & \\
$v_{sw}$     & $514 \pm 53$      & $518 \pm 39 $     &$ 532 \pm 38 $     & $697 \pm 96$ & {$\mathrm{km\,s^{-1}}$} \\
$n_a$        & $1.17 \pm 0.1 $   & $1.56 \pm 0.24 $  &$ 1.5 \pm 0.26 $   & $1.51 \pm 0.27$ & \\
$n_b$        & $1.64 \pm 0.25$   & $1.78 \pm 0.17$   & $1.56 \pm 0.26$   & $1.51 \pm 0.27$ & \\
N-RMSE ($\epsilon$) & $0.37 \pm 0.02$ &$ 0.27 \pm 0.01$ &$ 0.34 \pm 0.02$ &$ 0.38 \pm 0.02 $& \\
$H_{GH}$ & $-2.17 \pm 1.1$ & $-12.60 \pm 1.0$ & $-2.31 \pm 1.9$ & $\times$ & {$\times 10^{41}\,\mathrm{Mx^2}$} \\
\hline
$H_{L}$  & $-376.24 \pm 40$ & $-23.35 \pm 2.6$& $-5.01 \pm 0.8$ & $\times$ & {$\times 10^{41}\,\mathrm{Mx^2}$} \\
$\chi^2$ & 5.587 & 0.540 & 0.948 & 0.629 & \\
$p_{L}$ & -0.752 & -0.780 & 0.146 & >1.0 & \\
\hline
\end{tabular}
\tablefoot{We compute the magnetic helicity using geometry parameters (radius, length...) through the MC crossing by every spacecraft, and we consider the geometry detailed in Section \ref{ss:dataInSitu}. Helicity calculations inferred with the 3DCORE model ($H_{GH}$), along with the helicity and impact parameter computed with the Lundquist model ($H_L$ and $p_L$, resp.) and the fit quality ($\boldsymbol{\chi^2}$), are shown in the last three rows.}
\end{table*}
\begin{figure}[!h]
\centering
\includegraphics[width=0.7\hsize]{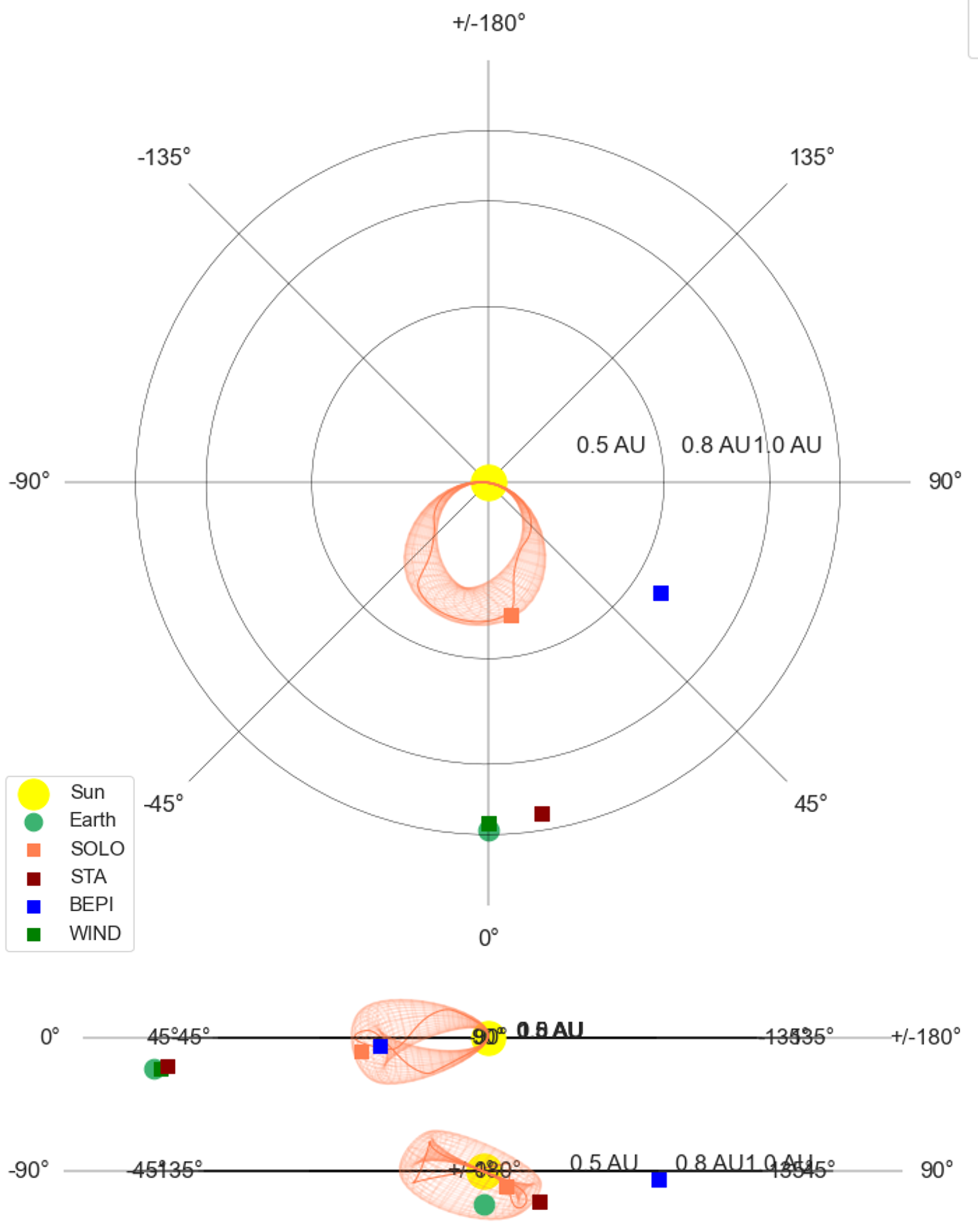}
    \caption{Representative magnetic flux rope reconstruction from SolO measurements at the time the MC was first measured in-situ by SolO (2024 March 23, 15:05:00 UT). The top panel presents a view from the solar equatorial north. Middle and bottom panels show a view from the $90^\circ$ and $0^\circ$ longitude along the Sun's equatorial plane, respectively. The solid line within the ICME structure represents a magnetic field line with the twist inferred from the model.}
        \label{fig:soloCME}
\end{figure} 

\begin{figure*}[!ht]
\centering
\includegraphics[width=0.75\hsize]{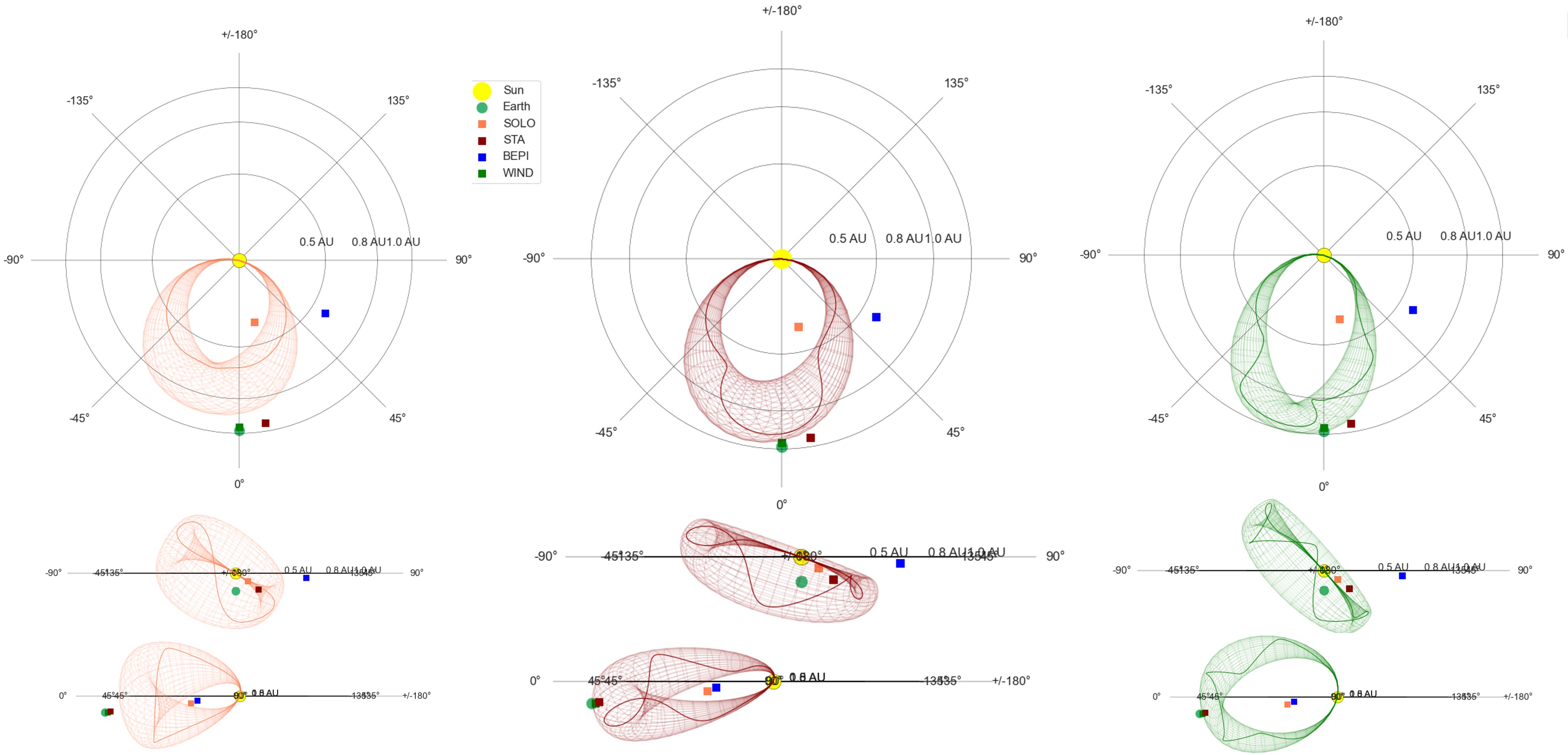}
    \caption{Reconstruction of the magnetic flux ropes (MFRs) based on SolO (left), STEREO-A (middle) and Wind (right) data. The SolO- and STEREO-based reconstructions were artificially advanced to the time when the MFR was detected by Wind on 2024 March 24 at 18:22 UT.}
        \label{fig:CME1au}
\end{figure*}

\indent Table \ref{t2} also lists the magnetic helicity inferred from 3DCORE with the geometrical assumptions (angular width and radius of the flux rope) we made (see Section \ref{ss:dataInSitu}). The first notable result is the difference in magnitude ($5$ to $10$ times higher) between the helicity derived from STEREO-A data and the other two spacecraft. This trend, along with the overall good consistency between SolO and Wind, suggests that STEREO-A is either probing a different part of the MFR, or the same MFR with a different angle originating from a kinked axis or a highly non-self-similar expansion. Regarding the impact of the geometry on the helicity calculations, it appears that the angular width change affects the results in different ways. For example, with SolO, the MFR enlargement (i.e., a larger angular width of the MFR twisted part considered for calculating the helicity) seems to reduce the measured helicity. On the contrary, with Wind, a larger angular width does not lead to a smaller value of helicity, but to minor differences between $30^\circ$ and $35^\circ$. The differences in cross-section between the MFR apex and the side of the twisted region have an impact on the calculated helicity. As we use these two cross-sections to quantify the uncertainties on the helicity estimate, we can study these differences and assess that the helicity values calculated with the apex radius are larger than those computed with the radius at the side of the considered twisted front. We also look at the impact of this twisted front region size by changing its angular width. If we link these results with the helicity budget at the AR, but also with the values found with a Lundquist fit, we can affirm that a $30^\circ$ MFR provides the most consistent results.\\
\indent Figure \ref{fig:soloCME} shows a representation of the MFR for SolO measurements at the time the start of the SolO MC was measured in-situ (i.e., 2024 March 23,15:05:00 UT). In the top panel, one can see a top view of the heliosphere up to the Earth distance, showing the longitudinal propagation direction of the ICME. The direction of propagation is slightly east-sided with respect to the Sun-Earth line, resulting in SolO probing the MFR a few degrees west of its apex. This is consistent with the observation made with coronographs (Figure \ref{fig:GCS} in Section \ref{ss:dataInSitu}) where the CME propagation seems to be directed a few degrees east of the Sun-Earth line. The two lower panels show views from the equatorial plane from a $90^\circ$ longitudinal angle (middle panel) and a $0^\circ$ one (bottom panel). These two views give a better perspective on the propagation direction latitude and on the tilt of the ICME, allowing us to learn more about its 3D structure. The 3DCORE reconstructions expanded at Wind detection time (2024 March 24 18:22 UT) in Figure \ref{fig:CME1au} display the 3D differences between the three spacecraft. By looking at them, and also at Figure \ref{fig:soloCME} and Figure \ref{fig:GCS}, we observe a similar inclination of $\sim60^\circ$ from the north axis. If we compare the inclination differences closely, we observe a smaller reconstructed tilt for STEREO-A and, at the same time, a larger tilt for Wind. Overall, this orientation of the CME is compatible with the H-$\alpha$ and EUV images, showing a Northeast-Southwest directed filament and hot post-flare loop structure (see Figure \ref{fig:corona}). The modeling also shows the twist by plotting a representative magnetic field line within the torus. This value is associated with the twist factor ($T_{fac}$ in Table \ref{t2}), which is negative for SolO, suggesting a left-handed MFR as described in Section \ref{ss:RemoteSensing}. Overall, these results confirm the general consistency among Solar Orbiter, STEREO-A, and Wind.\\
\indent By selecting an appropriate snapshot time, the reconstructions can be compared at 1 AU by forward modeling each of the fitted models to the desired time. Since Wind is the spacecraft located at L1, we adopt its MC detection time (2024 March 24, 18:22 UT) as the reference for this comparison, as shown in Figure \ref{fig:CME1au}. A notable discrepancy emerges for Solar Orbiter, whose drag-based propagation appears too slow to reproduce the observations time near 1 AU. This behavior is consistent with the differences in the inferred expansion rate ($n_a$ in Table \ref{t2}) between Solar Orbiter and the spacecraft situated close to 1 AU. Regarding the general structure of the ICME, we can notice a clear similarity in the field line representation between the first two panels (SolO and STEREO-A), determined by the twist factor parameter ($T_{fac}$ in Table \ref{t1} and \ref{t2}), which is higher for Wind (around $-20.8$ against $-48.8$). This behavior changes our view on the ICME propagation, as it supports the idea that SolO and STEREO-A probe the same part of the MFR. However, the uncertainty level does not allow us to confirm this fully. Moreover, the shape differences originate from the non-symmetrical structure of ICMEs, making them strongly sensitive to the spacecraft location, especially in a one-spacecraft reconstruction \citep[see][chap. ~4.5]{al-haddad_magnetic_2025}.
\section{Summary and Discussion}\label{s:SumDis}
We studied the magnetic helicity evolution of an ICME event erupting from an X1.1 class flare in AR 13614 (SOL-2024-03-23T00:58X1.1) on 2024 March 23. Our aim is to trace the magnetic helicity budget during the flare/CME event to the in-situ counterpart, with the magnetic field components being measured on the ICME path by different spacecraft and at different distances from the Sun. From the in-situ measurements provided by four different spacecraft in the heliosphere (SolO, Wind, STEREO-A and BepiColombo), we use two different methods based on different flux rope models (Lundquist and the Gold-Hoyle-like 3DCORE), both detailed in Section \ref{ss:dataInSitu}. The 3DCORE model was chosen as a more elaborate approach to infer helicity because of the magnetic model used (non-linear force-free) and because of the elliptical cross-section geometry it is based on. As with Lundquist, this model was fitted to the in-situ measurements to retrieve the ICME parameters and, thereby, provided 3D reconstructions of the ICME in the heliosphere. Ultimately, and as a comparison, the AR magnetic field was reconstructed from HMI vector magnetograms by using a physics-informed neural network and the helicity inferred from it using a finite-volume method.\\
\indent From the simpler approach of a linear force-free configuration (Lundquist), we find a helicity content of the MC in overall agreement with the helicity ejected from the solar corona during the flare/CME event ($-5.0 \pm 1.0\times 10^{41} \text{ Mx}^2 $ vs. $ -2.2\pm 1.3 \times 10^{41} \text{ Mx}^2$, resp.), when the impact parameter is sufficiently low ($p_L\simeq 0.2$ at Wind). A value of the impact parameter of $p_L\gtrsim 0.7$ indicates that a spacecraft (here Solar Orbiter and STEREO-A) probed the MC far away from the central axis, and hence that the corresponding estimates of physical parameters are of limited reliability. Therefore, the substantially larger values of MC helicity at lower radial distances (for SolO and STEREO-A, $\approx 10^{43} \text{ Mx}^2$ and $\approx 10^{42} \text{ Mx}^2$, resp.) cannot be interpreted in the sense of a radial decrease of the helicity content, e.g., the erosion of the MC in interplanetary space. These results need to be mitigated as the quality of the fit is low (see quality parameters in Table \ref{t2}).\\
\indent From 3DCORE, a more sophisticated approach of a uniformly twisted non-force-free field (Gold-Hoyle), we find a helicity content of the MC at SolO and Wind ($-2.2 \pm 1.1 \times 10^{41} \text{ Mx}^2$ and $-2.3 \pm 1.9 \times 10^{41} \text{ Mx}^2$, resp.) suggesting a fairly constant helicity content of the MFR during its propagation from the Sun to Earth. The STEREO-A-based estimate is a factor of about 5 times larger, which we attribute to the slightly different geometry of the 3DCORE MC reconstruction. It appears clearly on STEREO-A's panel in Figure \ref{fig:allfit} (middle row). Comparing this reconstruction to the other two, we notice the large cross-section of the MFR, especially compared to Wind (right panel), which is at a very similar radial distance (0.96 AU  for STEREO and 0.99 for Wind). This difference possibly comes from the local probing of the ICME. 3DCORE assumes an entirely coherent structure and a self-similar expansion; the results might differ from reality due to a kinked axis, non-uniform twist or deformations during the propagation. This problem was raised in previous studies such as \citet{zhang_studying_2025}, where they found only 36\% of events being self-similar expanding. These effects require multi-viewpoint observations to be studied in order to have a complete understanding of an event.\\
\indent When computing magnetic helicity from these model results, BepiColombo was excluded due to its position, the rotational characteristics of its measurements, and the quality of its reconstructions. BepiColombo's position and the ICME propagation direction suggest that the spacecraft sampled the flank of the magnetic flux rope, where the twist is supposed to be less concentrated compared to the apex region \citep{owens_legs_2016}, but the twist inferred with 3DCORE is larger at BepiColombo than at Wind or Solar Orbiter, for example. This contradiction may come from the fact that BepiColombo is probing the second ICME detailed in section \ref{ss:RemoteSensing}, especially since the handedness of its erupting flux rope is positive. However, the propagation direction of this second event is not in the direction of BepiColombo and thus does not allow us to conclude on this hypothesis.\\
\indent The remote sensing study of the source active region also provides helicity estimates in line with the one inferred in-situ. The helicity budget for the solar host AR was estimated using a finite-volume helicity method, applied to a time sequence of NLFF models based on a physics-informed neural network \citep[NF2;][]{jarolim_probing_2023}, and covering the pre- and post-flare time intervals. We find a flare/CME-related change in current-carrying helicity of $-2.16\pm1.32\times10^{41}~\text{Mx}^2$ (see Figure \ref{fig:HBudget}), which differs by at most a factor of 6 with respect to the in-situ-based ICME helicity estimates (see values listed in Table \ref{t2}).\\
\indent The overall consistency between our solar source-region and ICME helicities (ignoring any interaction processes) can be understood in an idealized scenario where the current-carrying part of the magnetic helicity describes the twisted magnetic core of the erupting structure and is expected to match the self-helicity of the ICME. This interpretation, however, is based on several assumptions and is challenged by the limitations of the measurements and models used in our study. One assumption is that we realistically capture the relevant coronal magnetic field structure with our NLFF models, including the flux rope later ejected and the surrounding magnetic field. In that respect, we ensured that the lateral and vertical extent of our extrapolation volume was appropriately selected. Another limitation of our approach is the possible transfer of current-carrying helicity to the volume-threading component (details in Section \ref{ss:RemoteSensing}), which can not be tested with NLFF modeling. Our method, considering the flare-related change of current helicity as a measure of the helicity injected into the erupting flux rope, remains physically motivated, as $H_J$ isolates the non-potential magnetic component, most directly associated with the ICME flux rope. On top of that, magnetic reconnection during and after the eruption can inject magnetic flux and helicity into the evolving flux rope and incorporate surrounding coronal structures, potentially increasing the helicity carried by the ICME. Previous studies have reported such helicity enhancement during flux-rope formation and evolution \citep[e.g.][]{qiu_magnetic_2007, hu_structures_2014, temmer_flare-cme_2017}, suggesting that the source-region helicity budget should be regarded as a lower limit to the helicity ultimately transported into interplanetary space. In addition, the in-situ spacecraft data provide only a 1D linear trajectory through a very large 3D ICME structure that is evolving in time. Fitting these profiles introduces significant geometric uncertainties and may strongly impact the inferred helicities (see parameters of Eq. \ref{eq:lundquist} and Eq. \ref{eq:GH}). In summary, the comparison between the remote-sensing (at the solar-source region) and in-situ-based (ICME) helicities can only quantitatively act as an order-of-magnitude constraint.
\section{Conclusion}
We present an analysis of an eruptive event that occurred early on 2024 March 23 in AR NOAA 13614. The event is characterized by an X1.1-class flare and the partial eruption of a filament, followed by the formation of a cusp-shaped coronal loop structure and a CME. We infer the magnetic helicity budget of the flare/CME event using data-constrained NLFF modeling (NF2) of the host AR and find a value of $-2.16\pm1.32\times10^{41}~\text{Mx}^2$. The interplanetary counterpart of the CME was probed by four different spacecraft at different radial distances between 0.39 AU and 0.99 AU. The in-situ magnetic field measured at different spacecraft locations during the ICME crossing was fitted using different underlying model approaches. One approach is based on a symmetric linear force-free configuration (Lundquist), and one is based on a uniformly twisted NLFF configuration (Gold-Hoyle). We find good consistency between the Gold-Hoyle-based estimates helicities at Wind and Solar Orbiter, and the helicity budget inferred at the AR. The single reliable Lundquist-based helicity estimate  -- at the Wind spacecraft position  -- is also very close to the amount of helicity inferred from the solar source region (difference between post- and pre-flare helicity), differing only by a factor of $\sim$two. The Gold-Hoyle-based ICME helicity estimates based on STEREO-A measurements, however, deviate to a larger degree (by a factor of $\sim$ 6) from the solar source region estimate. These differences are also visible on the 3D reconstructions of the ICME. In particular, the cross-sectional area seems larger, so that the radius of the reconstructed MFR is larger, which directly translates to a larger ICME helicity estimate. This finding implies that, even though the position of two spacecraft relative to the Sun is similar (as for STEREO-A and Wind in our study, separated by only $\sim$ 10 degrees), the ICME crossing can be substantially different -- considerably longer for STEREO-A in this study. Another way to interpret our STEREO-A-based ICME properties is that they represent a violation of the underlying assumption of a self-similar radial expansion of the ICME.\\
\indent In summary, our flare/CME helicity budget is a lower limit for the helicity shed by NOAA AR 13614. This helicity is then carried away with the ICME in the Heliosphere and appears, according to the more reliable estimates, to travel relatively undisturbed until 1 AU. It is to be noted, however, that the corresponding interpretation is challenging because of the different methods used. Similar analyses on a set of events or targeted at more complex cases yet represent another challenge and are anticipated for the future.
\begin{acknowledgements}
      This work was funded in parts by the Austrian Science Fund (FWF) 10.55776/PAT7894023. This work is supported by ERC grant (HELIO4CAST, 10.3030/101042188) funded by the European Union. M.D. acknowledges support from EU project SPEARHEAD (GA 101135044), within the scope of which the FDAT tool was developed (https://github.com/spearhead-he/FDAT). Views and opinions expressed are, however, those of the author(s) only and do not necessarily reflect those of the European Union or the European Health and Digital Executive Agency (HaDEA). Neither the European Union nor the granting authority can be held responsible for them. A.M., J.K.T., A.M.V., and M.D. acknowledge the support from the Austrian–Croatian Bilateral Scientific Project “Analysis of solar eruptive phenomena from cradle to grave”. We acknowledge the Community Coordinated Modeling Center (CCMC) at Goddard Space Flight Center for the use of the DONKI system, \url{https://kauai.ccmc.gsfc.nasa.gov/DONKI/}.
\end{acknowledgements}
\bibliographystyle{bibtex/aa.bst}
\bibliography{bibtex/bib.bib}
\begin{appendix}
\onecolumn
\section{The case of BepiColombo}\label{ap:BEPIresults}
\begin{figure*}[ht]
    \centering
    \includegraphics[width=0.65\linewidth]{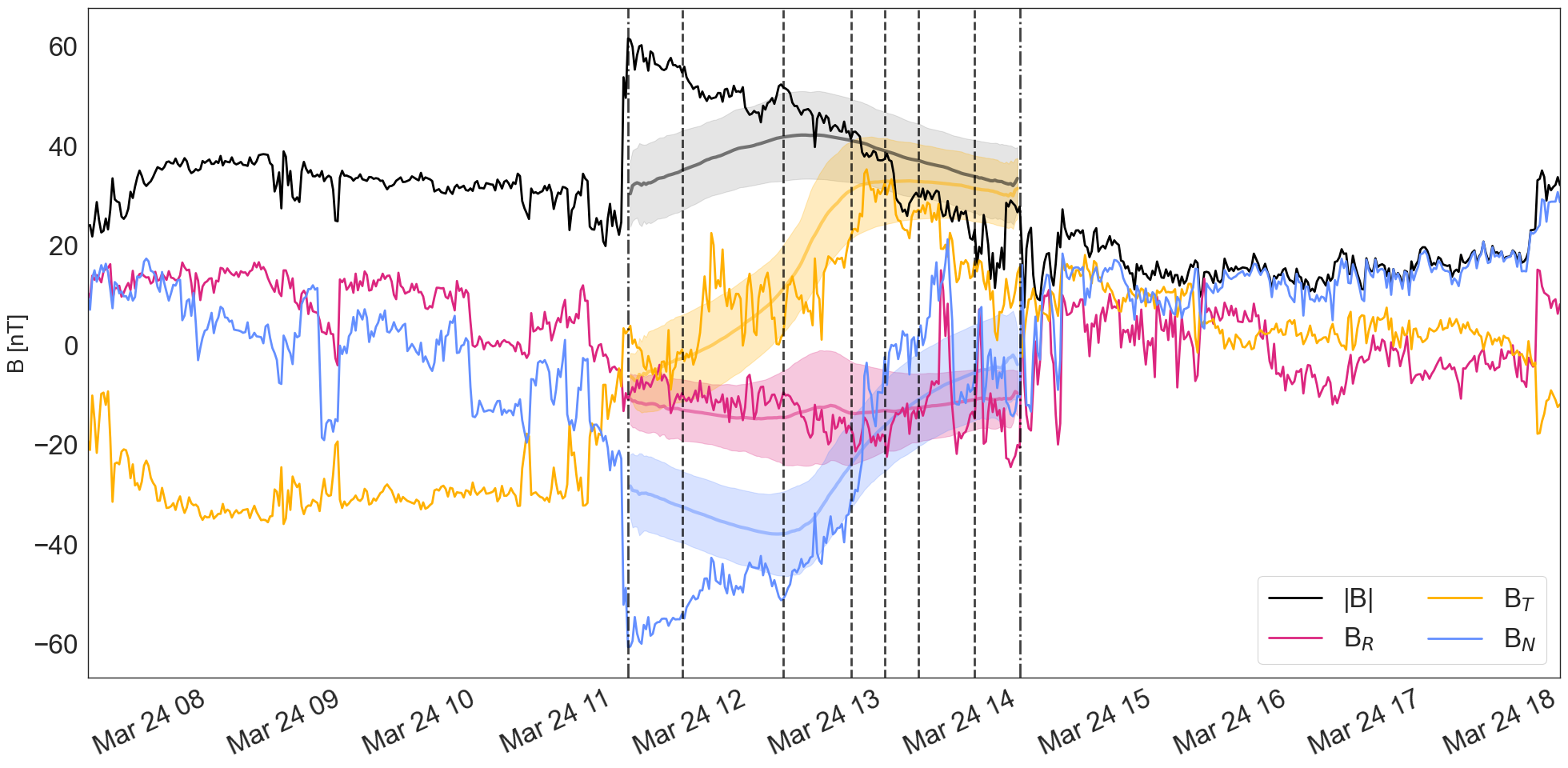}
    \includegraphics[width=0.30\linewidth]{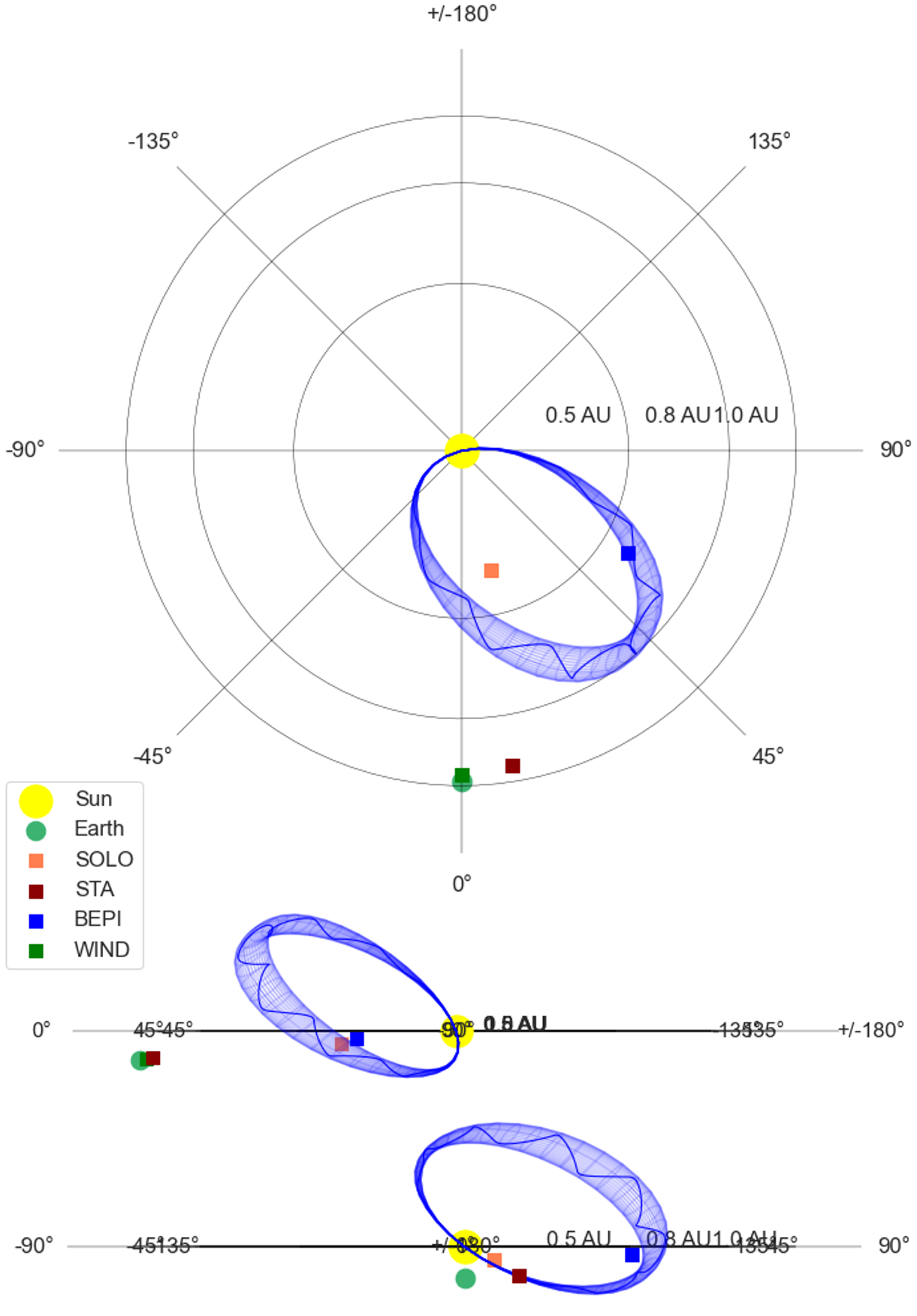}
    \caption{3DCORE model results for BepiColombo's magnetic field measurements. The left panel shows the fitted profiles as in Figure \ref{fig:allfit} and the right one shows the reconstruction of the CME at BepiColombo's detection time (2024-03-24 11:06 UT).}
        \label{fig:LundAnnex}
\end{figure*}

Out of the four spacecraft that observed the studied ICME event, BepiColombo (hereafter only Bepi) is the one showing very different characteristics. In particular, Bepi's magnetic field measurements of the MC portray a different type of MFR than the one detected by every other spacecraft. The value of the tangential component Bt remains mostly positive throughout the MC, and therefore does not match the other spacecraft measurements where this component remains mostly negative. Additionally, Bepi's measurements do not show a clear shock and sheath region ahead of the MC (see Bepi's panel in Figure \ref{fig:allInSitu}), which can suggest a slower propagation in Bepi's direction, preventing the compression of the solar wind. By looking at the event's propagation direction, it appears clear that Bepi is probing the ICME structure through its leg. Despite this situation, and to ensure that our interpretation is satisfactory, every model is performed on Bepi's data, and the results are shown in Figure \ref{fig:LundAnnex}.\\
Looking at the left panel, which shows the 3DCORE fitting of the MC measurements (details in Section \ref{ss:dataInSitu}), we notice that this model fits the signature quite well, with only a small deviation towards the start of the MC. The 3D reconstruction from this fit (right panel) does not reconcile with the reconstruction at the other spacecraft in our constellation. Notably, the propagation direction parameters (longitude, latitude, inclination in Table \ref{t2}) at Bepi are all in strong disagreement with the parameters at other spacecraft. Moreover, the twist factor ($T_{fac}$ in the same table) is not only higher for Bepi, but also has a different sign, meaning a different handedness of the MFR. This result could lead to thinking that Bepi is probing a different event, but this interpretation would not explain the absence of a shock. This high twist factor seems also quite nonphysical, as described in \citet{al-haddad_magnetic_2025} and \citep{torok_sun--earth_2018}. Looking at the 3D reconstruction, the ICME inclination is low, but the direction of propagation is very different from the reconstruction at any other spacecraft (see Figure \ref{fig:CME1au}). Finally, the Lundquist fit is not able to fit the in-situ measurements, with a final impact parameter of $\sim$1. It even provides a better fit by choosing a positive handedness as an initial guess, which is contradictory to every other spacecraft measurement and to the remote sensing analysis of the source active region. This characteristic is in line with the positive $T_{fac}$ inferred with 3DCORE, which again can lead us to think that Bepi is probing a different event or the leg of the studied one. These results lead us to not consider the helicity content inferred from this data as a reliable estimate for the studied ICME.
\end{appendix}
\end{document}